%% file: main.tex
\documentclass[twocolumn,tighten]{aastex62}
\hypersetup{linkcolor=red,citecolor=blue,filecolor=cyan,urlcolor=magenta}
\usepackage{amsmath}
\usepackage{natbib}
\usepackage{color}
\usepackage{hyperref}
\usepackage{url}

\input{newcommands.tex} 
\newcommand{\pen}{\mathrm{penalty}}

\begin{document}

\title{
Dynamical clock of the Helmi stream---Analysis of the clumping of stars in the orbital frequency-space
}

\shorttitle{Dynamical age of the Helmi stream}
\shortauthors{K.~Hattori}

\author[0000-0001-6924-8862]{Kohei~Hattori}
\affiliation{National Astronomical Observatory of Japan, 2-21-1 Osawa, Mitaka, Tokyo 181-8588, Japan}
\affiliation{The Graduate University for Advanced Studies, SOKENDAI, 2-21-1 Osawa, Mitaka, Tokyo 181-8588, Japan}
\affiliation{The Institute of Statistical Mathematics, 10-3 Midoricho, Tachikawa, Tokyo 190-8562, Japan}
\email{Email:\ kohei.hattori@nao.c.jp}

\begin{abstract}

Reconstructing the assembly history of the Milky Way requires precise constraints on the dynamical age of its merger remnants—the time elapsed since a progenitor satellite was disrupted by the Galactic tidal force. We present a new framework to derive this dynamical age for disrupted stellar systems by extending the Fourier analysis of the orbital frequency distribution proposed by G\'omez \& Helmi. To overcome the smearing of frequency-space structures caused by observational noise, we introduce the Greedy Optimistic Clustering algorithm. This method allows for an optimistic exploration of the density contrasts in the orbital frequency space by taking into account the observational uncertainty in the data, effectively sharpening the signal required for age estimation. By applying this method to the Helmi stream, we derive a dynamical age of $6.8 \pm 0.8$ Gyr. Our derived accretion epoch is consistent with the observed kinematic properties of the Helmi stream. In particular, the marked asymmetry in the vertical velocity distribution---where approximately two-thirds of the stars have negative $v_z$ in the solar neighborhood---supports a relatively recent arrival. This suggests that the progenitor of the Helmi stream was accreted during an epoch of Galactic growth distinct from the much earlier Gaia-Sausage-Enceladus merger ($\sim 10$ Gyr ago). We validate our methodology using error-added mock simulations, demonstrating the reliability of our approach. Our results establish the Greedy Optimistic Clustering framework as a powerful chronometric tool for reconstructing the hierarchical assembly of the Milky Way using current and future high-precision astrometric datasets.

\end{abstract}


\section{Introduction}

\subsection{Chronology of the Milky Way formation}

The standard $\Lambda$CDM cosmology predicts that the Milky Way was assembled through hierarchical mergers of smaller stellar systems, such as dwarf galaxies and globular clusters. Indeed, recent stellar surveys have revealed the relics of these accretion events in the form of stellar streams or kinematic substructures in the Galactic halo \citep{Helmi1999Natur.402...53H, Belokurov2006ApJ...642L.137B, Helmi2008A&ARv..15..145H, Helmi2020ARAA..58..205H, Ibata2019ApJ...872..152I, Ibata2021ApJ...914..123I, Ibata2024ApJ...967...89I, Vasiliev2023Galax..11...59V, Bonaca2025NewAR.10001713B}. 

While the spatial distribution, kinematics, and chemical abundances of halo stars are now measured with unprecedented precision thanks to the Gaia mission and complementary spectroscopic surveys, the timing of individual accretion events remains elusive. Determining the dynamical age of each stellar stream or substructure is essential for reconstructing the chronological sequence of the Milky Way's formation \citep{McMillan2008MNRAS.390..429M, Gomez2010MNRAS.401.2285G, Li2026AJ....171..160L}. This temporal information complements chemical and orbital diagnostics, offering a more complete picture of the Galaxy's hierarchical assembly. By constraining when each merging event took place, we can better understand the growth history of the Galactic halo.

\subsection{The Helmi stream and its accretion epoch}

The Helmi stream is one of the earliest identified substructures in the Galactic halo, discovered by \citet{Helmi1999Natur.402...53H} through analysis of Hipparcos data of metal-poor stars \citep{Beers1995ApJS...96..175B, Perryman1997A&A...323L..49P, Chiba1998AJ....115..168C}. It is characterized by a prograde motion and a highly inclined orbit with respect to the Galactic plane. A distinctive feature of the stream is the asymmetry in its vertical velocity distribution, with a predominance of stars moving with negative vertical velocity ($v_z < 0$) \citep{Helmi1999Natur.402...53H, Kepley2007AJ....134.1579K}. The stream contains some relatively metal-rich stars (up to [Fe/H]$\simeq -1.3$), suggesting that its progenitor system was a massive dwarf galaxy \citep{Koppelman2019A&A...625A...5K}. These properties make the Helmi stream a benchmark for studies of stellar streams and accretion remnants, offering valuable insights into the early stages of the Milky Way's formation.

Several studies have attempted to estimate the accretion epoch of the Helmi stream using indirect methods. \citet{Koppelman2019A&A...625A...5K} performed $N$-body simulations of satellite disruption and found that an accretion time of 5--8 Gyr ago, along with a progenitor mass of $\sim 10^8 \msun$, reproduces the observed kinematic asymmetry in the vertical velocity distribution. This result supports earlier findings by \citet{Kepley2007AJ....134.1579K}. \citet{Ruiz-Lara2022A&A...668L..10R} analyzed the color-magnitude diagram of Helmi stream stars and inferred a quenching of star formation around 8 Gyr ago, interpreted as the time of accretion. More recently, \citet{Lindsay2025ApJ...989..189L} used asteroseismic data of two bright stars likely associated with the stream, finding that their ages exceed $\sim 11$ Gyr. These ages provide an upper limit on the accretion time.

Despite the Helmi stream's importance and the variety of approaches used to study its origin, there has been no direct measurement of its dynamical age based purely on its present-day phase-space structure. Existing estimates rely on indirect indicators such as stellar ages, star-formation histories, or comparisons with simulated disruption models. A direct dynamical age determination would offer a crucial constraint on the stream's accretion history and provide a critical insight into the formation history of the Milky Way.

\subsection{\GH\ method of deriving the stellar stream's dynamical age}

A powerful method for estimating the dynamical age of a disrupted stellar system was proposed by \cite{Gomez2010MNRAS.401.2285G}. When a satellite galaxy is tidally disrupted by the Milky Way, its stars disperse along their orbits but retain coherent structures in the phase space of orbital properties, such as the orbital actions $\vector{J}$ or orbital frequencies $\vector{\Omega} = (\Omega_R, \Omega_\phi, \Omega_z)$. In orbital-frequency space, these stars typically form a single, relatively large blob, reflecting their similar values of $\vector{\Omega}$ while exhibiting some intrinsic dispersion. However, when one focuses on a subset of these stars located in the solar neighborhood (or within any small spatial volume in the Galaxy), the distribution reveals a distinct pattern, as first demonstrated by \cite{McMillan2008MNRAS.390..429M}. Stars located far from the Sun are absent from this subset, meaning that the local selection effectively imposes a mask based on the current orbital phase. Numerical simulations show that, under such selection, the frequency-space distribution develops a semi-regular lattice of narrow clumps in, for example, the $(\Omega_R, \Omega_\phi)$ plane. Each clump, embedded within the broader blob, corresponds to stars that have completed a different integer number of orbital revolutions since the disruption event. For example, one clump may correspond to stars that have orbited the Milky Way $n$ times, while the adjacent clump corresponds to those that have orbited it $n+1$ times. The spacing between adjacent clumps, $\delta \Omega$, is set by the time elapsed since the disruption and follows the relation $\delta \Omega \simeq 2\pi / T_\mathrm{accretion}$. Following the method proposed by \cite{Gomez2010MNRAS.401.2285G}, this characteristic spacing can be measured via a two-dimensional Fourier transform of the frequency-space distribution, from which one can directly infer the disruption time. This approach is particularly appealing because it links observable structures to the dynamical age without requiring a full $N$-body simulation, though it remains dependent on the assumed Milky Way gravitational potential used to compute the orbital frequencies.

In practice, however, the application of this method requires high-precision phase-space information. In real observational data, uncertainties in positions and velocities propagate into the computed orbital frequencies, blurring the clump pattern in frequency space. This blurring can obscure the regular lattice structure and reduce the effectiveness of Fourier-based techniques for identifying the characteristic spacing $\delta \Omega$. Therefore, while the method proposed by \citet{Gomez2010MNRAS.401.2285G} offers a conceptually elegant and direct route to estimating the dynamical age, its implementation must carefully account for observational uncertainties.

\subsection{Scope of this paper}

In this paper, we tackle a key limitation of the method proposed by \citet{Gomez2010MNRAS.401.2285G}—its sensitivity to observational uncertainties—by employing a novel clustering technique called the Greedy Optimistic Clustering algorithm \citep{OkunoHattori2025,Hattori2023ApJ...946...48H}. This algorithm is specifically designed to identify clumps in a data set blurred by observational errors, which is ideal for resolving the underlying discrete clumps in orbital-frequency space that are necessary for dynamical age estimation. By applying this framework---which we validate through test-particle simulations---to the Helmi stream stars, we provide the first direct estimate of the Helmi stream's accretion epoch based on frequency-space clustering.

This paper is organized as follows. 
Section~\ref{sec:showcase} provides the conceptual overview of the method of \cite{Gomez2010MNRAS.401.2285G}. 
Section~\ref{sec:observation} presents the Gaia DR3 observational data used in our study. 
Section~\ref{sec:uncertainty_set} explains the construction of the uncertainty sets, which serve as the input for the Greedy Optimistic Clustering algorithm. 
Section~\ref{sec:GOEM_freq} details the Greedy Optimistic Clustering analysis, while Section~\ref{sec:FT_analysis} describes the Fourier analysis used to derive our primary result: the dynamical age of the Helmi stream. 
To ensure the reliability of our new framework, Section~\ref{sec:validation} provides an extensive validation using error-added mock simulations. 
Finally, Section~\ref{sec:discussion} discusses the implications of our results and provides our conclusions.

\begin{figure*}
\centering 
\includegraphics[width=0.27\textwidth ]{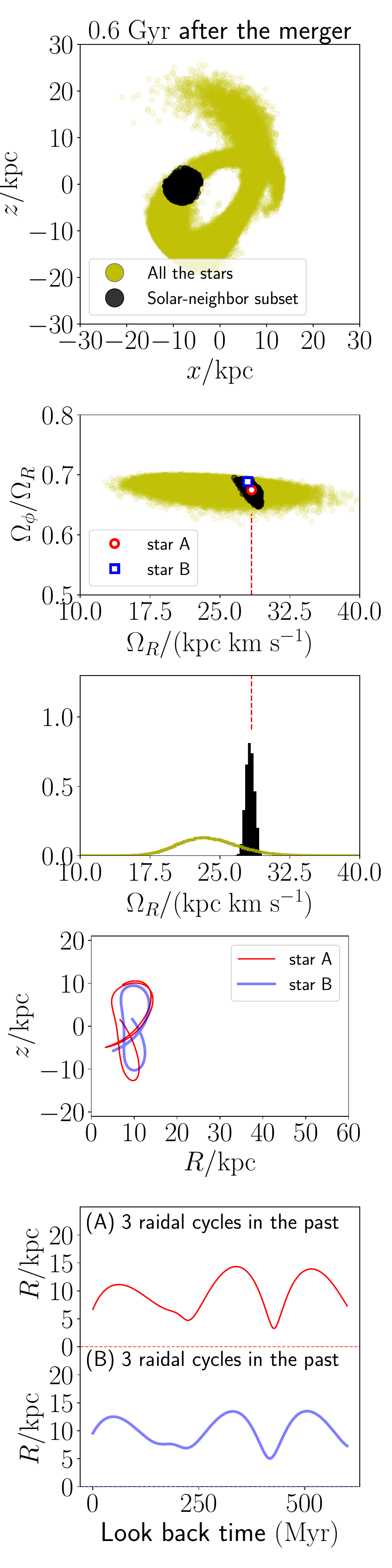}
\includegraphics[width=0.27\textwidth ]{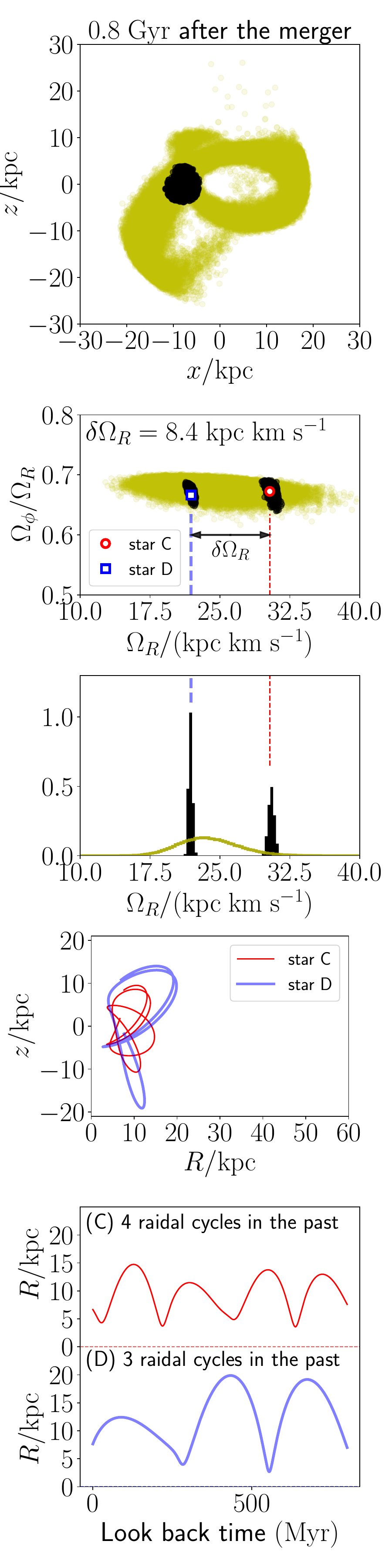}
\includegraphics[width=0.27\textwidth ]{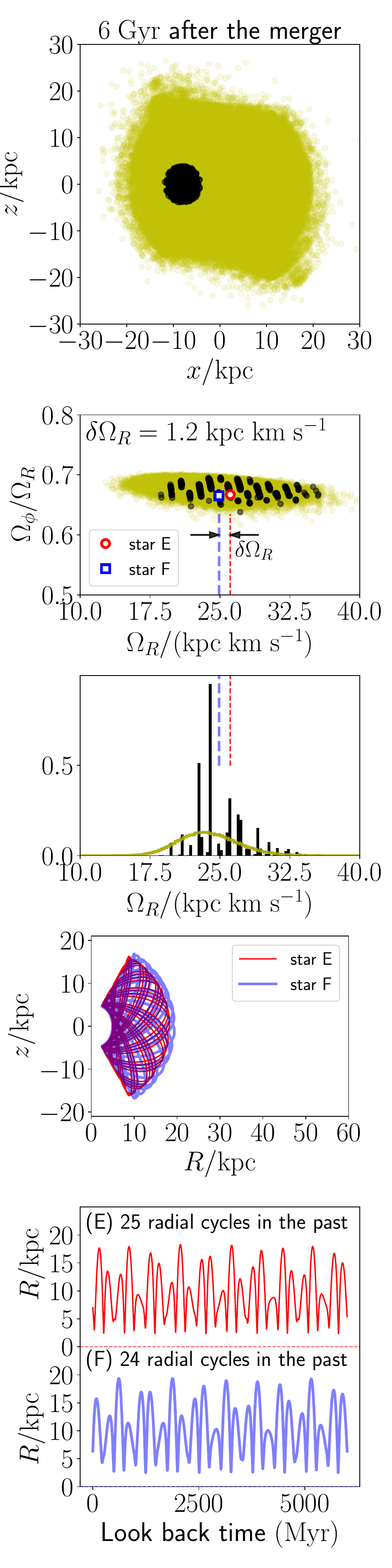}
\caption{
Test-particle simulation of a Helmi-stream-like disruption that occurred $6\ \Gyr$ ago. 
The left, middle, and right columns show snapshots taken $0.6$, $0.8$, and $6$ Gyr after the onset of accretion (the right column corresponds to the present epoch). 
The top and second rows display the spatial and orbital-frequency distributions.
Yellow dots mark all stream stars, and black dots indicate a subset of stars inside the solar-neighbor volume defined by the current-day location of the Sun. 
As shown in the second row, the solar-neighbor subset develops a semi-regular lattice of clumps in frequency space, with the spacing in $\Omega_R$ ($\delta\Omega$) decreasing over time.
In the third row, black and yellow histograms indicate 
the distribution of $\omr$ for all the stars and the solar-neighbor subset, respectively. 
For each snapshot, two representative stars in the solar-neighbor subset (red circle, blue square) are highlighted---stars A and B at $0.6$ Gyr, C and D at $0.8$ Gyr, and E and F at $6$ Gyr. 
Their orbits in the meridional plane and the time variation of $R$ are shown in the fourth and fifth rows. 
Stars in adjacent clumps differ by exactly one radial cycle (e.g., star E has completed 25 while star F has 24).
For details, see the text.
}
\label{fig:simulation_no_error}
\end{figure*}

\section{Conceptual overview of the \GH\ Method}
\label{sec:showcase}

Before turning to the analysis of the observational data, we first summarize the
conceptual basis of the \GH\ method using idealized, noise-free examples.

When a satellite galaxy is accreted by the Milky Way, it becomes tidally disrupted by the Galactic tidal force, releasing its stars along the progenitor's orbit \citep{Johnston1996ApJ...465..278J, Binney2008MNRAS.386L..47B, Koposov2010ApJ...712..260K}. Although the debris eventually spreads over a large volume in configuration space, the orbits of the stars remain dynamically coherent and can be compactly described in terms of action-angle variables \citep{Eyre2011MNRAS.413.1852E, Sanders2013MNRAS.433.1813S, Sanders2013MNRAS.433.1826S, Bovy2014ApJ...795...95B}. If the Galactic potential is static, each star is characterized by a set of orbital frequencies $(\Omega_R, \Omega_\phi, \Omega_z)$ that remain constant over time, providing a powerful tool for tracing the dynamical evolution of disrupted systems.

To illustrate the underlying idea of the method of \cite{Gomez2010MNRAS.401.2285G}, we present a noise-free test-particle simulation of a Helmi-stream-like system whose tidal disruption began $6$ Gyr ago within a static Galactic potential (Fig.~\ref{fig:simulation_no_error}).
The left, middle, and right columns of the figure show snapshots taken $0.6$, $0.8$, and $6\ \mathrm{Gyr}$ after the onset of accretion, respectively. The top and second rows display the spatial and orbital-frequency distributions. Yellow dots mark all stars originating from the progenitor, while black dots indicate a subset located within the solar-neighbor volume defined by the current-day position of the Sun.
Here, we note that the solar-neighbor subset is defined separately at each snapshot. Specifically, we define the `solar-neighbor volume' as a sphere with a radius of 4 kpc centered at $(x,y,z)=(-8.277, 0, 0.0208)$ kpc, and this definition is fixed for all epochs. Consequently, the set of stars lying inside this volume (i.e., the solar-neighbor subset) changes with time.

As the disruption proceeds, the debris spreads out in configuration space, yet in frequency space the solar-neighbor subset forms a semi-regular lattice of narrow clumps. Each clump corresponds to stars that have completed an integer number of radial oscillations since the onset of accretion. The spacing between neighboring clumps in $\Omega_R$ encodes the elapsed time since disruption: two wraps differing by one full radial cycle have frequency separation
\begin{equation}
\delta \Omega = \frac{2\pi}{T_\mathrm{accretion}}, 
\end{equation}
where $T_\mathrm{accretion}$ is the time since the onset of accretion. Consequently, if the clumps are more densely distributed in frequency space---that is, if the spacing $\delta\Omega$ is smaller---it implies that the disruption occurred further in the past.\footnote{
Suppose that two adjacent clumps have radial frequencies of $\Omega_R$ and $\Omega_R - \delta\Omega$.
If the stars with radial frequency $\Omega_R$ have completed $n_R$ radial oscillations since the onset of accretion,
the stars with frequency $\Omega_R - \delta\Omega$ have completed $(n_R-1)$ oscillations.
Because one full radial cycle corresponds to a period $2\pi/\Omega_R$, the total elapsed time $T_\mathrm{accretion}$ satisfies $T_\mathrm{accretion}=n_R \times 2\pi/\Omega_R = (n_R-1) \times 2\pi/(\Omega_R - \delta\Omega)$. 
Eliminating $n_R$ from these expressions yields 
$T_\mathrm{accretion}=2\pi/\delta\Omega$, 
which shows that the spacing in radial frequency directly encodes the dynamical age of the stream. The same argument holds for the azimuthal frequency. 
}

The third row of Fig.~\ref{fig:simulation_no_error} shows histograms of $\Omega_R$ for all stars (yellow) and for the solar-neighbor subset (black). 
At the snapshot $0.6\ \Gyr$ after the onset of accretion, only a single wrap of the stream is present within the solar-neighbor volume, resulting in a single peak in the $\Omega_R$ histogram. 
The solar-neighbor stars~A and~B, randomly selected from this clump, have completed three radial oscillations since the accretion. 

At $0.8\ \Gyr$, two wraps of the stream are visible, producing a double-peaked $\Omega_R$ distribution with a spacing of $\delta\Omega = 8.4\ \kpc\ \kms$. 
The solar-neighbor stars~C and~D, taken from these adjacent clumps, have completed four and three radial oscillations, respectively. 
From the spacing, the dynamical age of the stream at this epoch is inferred to be 
$T_\mathrm{accretion} = 2\pi / \delta\Omega = 0.75\ \Gyr$, in good agreement with the true value of $0.8\ \Gyr$. 

By $6\ \Gyr$ after accretion, many wraps of the stream overlap in the solar-neighbor volume, yielding multiple peaks in the $\Omega_R$ histogram with a typical spacing of $\delta\Omega = 1.2\ \kpc\ \kms$. 
The solar-neighbor stars~E and~F, taken from two adjacent clumps, have completed 25 and 24 radial oscillations since the accretion. 
The corresponding dynamical age, $T_\mathrm{accretion} = 2\pi / \delta\Omega = 5.2\ \Gyr$, again closely matches the true elapsed time of $6\ \Gyr$.

The fourth and fifth rows of Fig.~\ref{fig:simulation_no_error} further illustrate this interpretation. 
The selected stars~C--F highlight that adjacent clumps differ by exactly one additional radial oscillation: 
their orbits in the meridional $(R, z)$ plane (fourth row) and their time variations of the Galactocentric radius $R(t)$ (fifth row) 
clearly show that, at the snapshots $0.8\ \Gyr$ and $6\ \Gyr$ after the onset of accretion, one star completes one more radial cycle than the other. 
These panels provide an intuitive visualization of how the spacing $\delta\Omega$ encodes the number of completed oscillations and, consequently, the dynamical age of the stream.

Quantitatively, \citet{Gomez2010MNRAS.401.2285G} proposed to measure this characteristic spacing via a two-dimensional Fourier transform of the stellar distribution in frequency space. The transformation converts the semi-regular lattice into a power spectrum whose dominant peak corresponds to $\delta \Omega$. For an idealized, noise-free dataset such as our simulation, the spectrum exhibits a sharp, well-defined peak, and the corresponding wavelength directly yields the disruption time through $T_\mathrm{accretion} = 2\pi / \delta \Omega$. The Fourier transform thus offers an objective way to infer the dynamical age of a stellar stream from its observed frequency distribution.

However, the success of the method by \citet{Gomez2010MNRAS.401.2285G} relies critically on accurate orbital-frequency measurements for each star. Even modest observational uncertainties in position or velocity smear the frequency distribution, wash out the clump pattern, and suppress the dominant Fourier peak. When Gaia-like observational error is added to the simulated data, the lattice structure largely disappears and the periodic signal becomes indiscernible. This limitation motivates the development of a denoising procedure that can recover the intrinsic structure from uncertain data, a task accomplished by the Greedy Optimistic Clustering method introduced in the following sections.

\begin{figure*}
\centering 
\includegraphics[width=0.95\textwidth ]{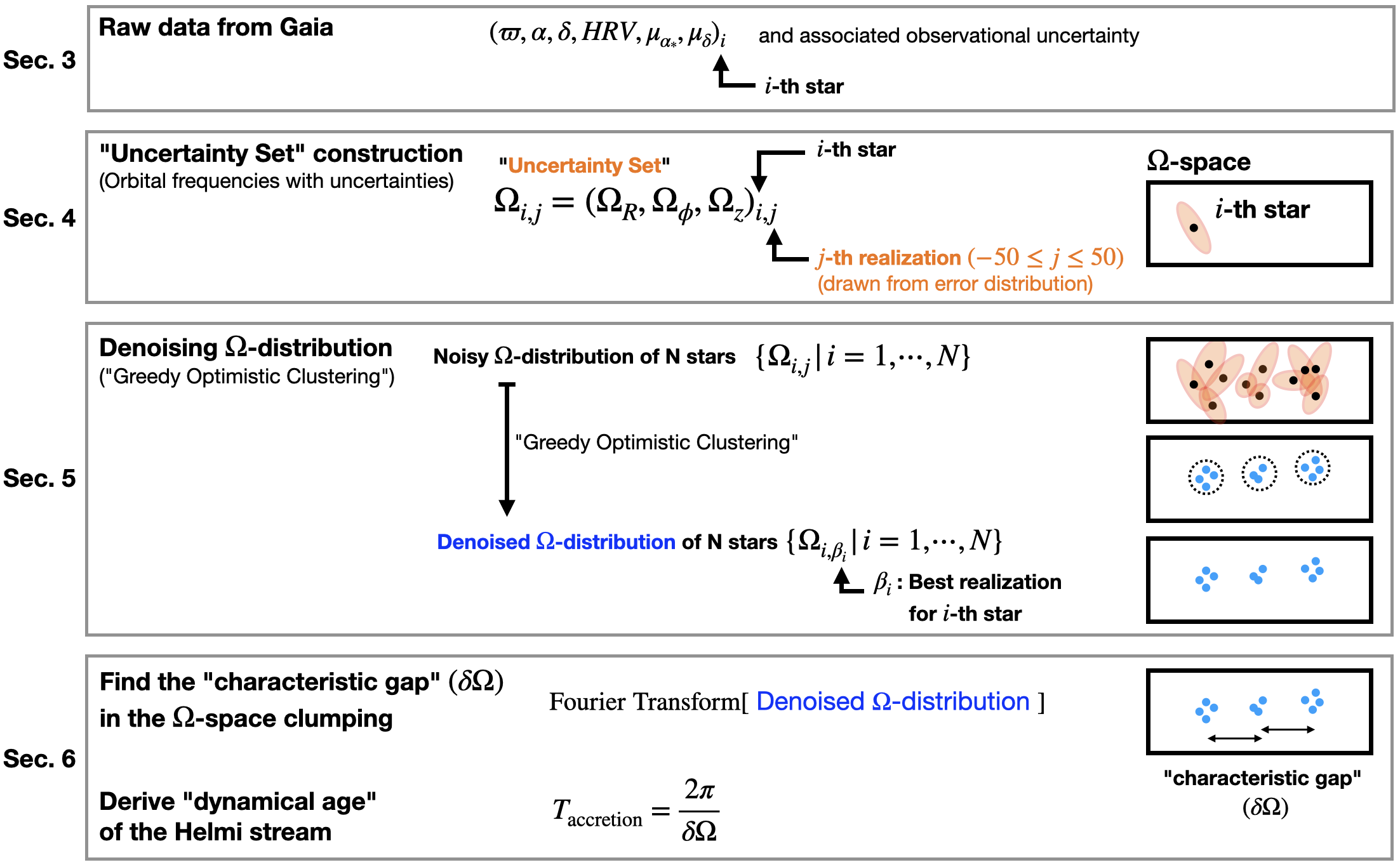}
\caption{
Schematic overview of the analysis pipeline for determining the dynamical age of the Helmi stream. The workflow follows the sequence within the paper. (Section 3) We start from the raw data of Helmi-stream stars selected from the Gaia data (positions, velocities, and associated uncertainties). (Section 4) For each star ($i$th star), we sample position and velocity from the error distribution and convert them to orbital frequency, generating the uncertainty set $\{ \Omega_{i,j} \mid j=-50, \dots, 50 \}$ consisting of 101 synthesized orbital-frequency vectors. Since the distance uncertainty dominates for most stars, the uncertainty set typically shows an elongated distribution, as illustrated by the orange ellipse in the inset (see also the middle column in Fig.~\ref{fig:summary_denoise_all}). (Section 5) We apply the Greedy Optimistic Clustering algorithm to these uncertainty sets for $N$ stars, to resolve discrete clumps in the $\Omega$-space. This procedure is technically a denoising process, and we use the resulting best configuration as the input for further analysis. (Section 6) We perform a Fourier analysis on the resulting denoised $\Omega$-distribution to identify the characteristic frequency gap, $\delta\Omega$. As illustrated in Section~\ref{sec:showcase} (see also Fig.~\ref{fig:simulation_no_error}), this gap is physically linked to the time elapsed since the progenitor's disruption, $T_{\mathrm{accretion}} = 2\pi / \delta\Omega$. While this schematic illustrates the logic of a single instance, our final result is derived from an ensemble of 400 independent denoising realizations to ensure a reliable estimate of $T_{\mathrm{accretion}}$. 
}
\label{fig:schematic}
\end{figure*}

\section{Observational Data} \label{sec:observation}

As outlined in Fig.~\ref{fig:schematic}, Sections~\ref{sec:observation}--\ref{sec:FT_analysis} describe the sequence of analyses used to derive the dynamical age of the Helmi stream. 
In this section, we begin by selecting candidate members of the Helmi stream  using position and velocity information from the Gaia DR3 catalog \citep{Gaia2021A&A...649A...1G, GaiaCollaboration2023A&A...674A...1G}, together with metallicity estimates taken from \citet{Andrae2023ApJS..267....8A}.

\subsection{Construction of a catalog of nearby stars with 6D position/velocity data and metallicity} \label{sec:sample_selection}

First, we crossmatch the Gaia DR3 catalog and the chemical catalog of 175 million stars in \cite{Andrae2023ApJS..267....8A}. 
For each star, we extract the Right Ascension and Declination $(\alpha, \delta)$, parallax $\varpi$, proper motion $(\mu_\alpha, \mu_\delta)$, and line-of-sight velocity $\vlos$, along with their associated uncertainties, from Gaia DR3. 
Throughout this paper, we adopt the parallax values corrected for the global zero-point offset, rather than using the raw measurements. The correction is applied using the publicly available code \texttt{gaiadr3-zeropoint}, and the corrected parallax is denoted as $\varpi$. 
From the catalog in \cite{Andrae2023ApJS..267....8A}, we extract the metallicity [M/H] (\texttt{mh\_xgboost} in their catalog), which has been derived from the Gaia XP spectra 
\citep{DeAngeli2023A&A...674A...2D, Montegriffo2023A&A...674A...3M, 
Gaia2016A&A...595A...1G, 
Gaia2018A&A...616A...1G, 
Gaia2021A&A...649A...1G, 
GaiaCollaboration2023A&A...674A...1G}.\footnote{While \citet{Hattori2025ApJ...980...90H} also provided a chemical catalog from Gaia XP spectra, that catalog is optimized for low dust extinction regions (e.g., $|b| > 20^\circ$). To maximize our sample size, we adopt the abundances from \citet{Andrae2023ApJS..267....8A} in this work.}

To select nearby stars with relatively high-quality astrometric data, we apply the following additional criteria: 
(i) \texttt{parallax\_over\_error}$>2$; 
(ii) $1/(\varpi + 2 \sigma_\varpi) <4 \kpc$; and 
(iii) \texttt{ruwe}$<1.4$. 
Here $\sigma_\varpi$ is the uncertainty in the parallax. 
We note that the quantity $1/(\varpi + 2 \sigma_\varpi)$ is the two-sigma level lower limit on the heliocentric distance and the conditions (i) and (ii) are designed to select solar-neighbor stars. 
To minimize contamination from the stellar disk, we further impose a metallicity cut of [M/H] $< -1$. 
Additionally, to avoid contamination from the Large and Small Magellanic Clouds (LMC and SMC), we exclude stars located within $10^\circ$ and $5^\circ$ of the LMC and SMC, respectively.

\subsection{Selection of candidates of stars associated with Helmi stream} \label{sec:sample_selection}

To identify possible member stars of the Helmi stream, we define a selection box (hereafter Box B, following the convention in \citealt{Koppelman2019A&A...625A...5K}) in the space of angular momenta.
The box is defined by
$750 < J_\phi / (\kpc \;\kms) < 1700$, 
$1600 < L_\perp / (\kpc \;\kms) < 3200$,
where $J_\phi = -L_z$ and $L_\perp = (L_x^2 + L_y^2)^{1/2}$. 

For each star in the subset of Gaia DR3 stars selected earlier, we vary its parallax within the $\pm 2 \sigma_\varpi$ range. 
Then we retain those stars that enter Box B for at least one parallax value within its 2$\sigma_\varpi$ uncertainty range that satisfies the Box B criterion. 
Also, by adopting the Galactic potential model in \cite{McMillan2017}, 
we remove one star that is likely unbound to the Milky Way. 
After these procedures, we obtained 
783 stars that are likely associated with the Helmi stream.

We note that 66\% of these stars (518 stars) satisfy $v_z<0$, 
which is consistent with previous works such as \cite{Koppelman2019A&A...625A...5K}. 
In order to minimize the contamination from non-Helmi stream stars, 
we use these 518 stars (with $v_z<0$) 
as our main sample to be analyzed in this paper.

\section{Construction of the Uncertainty Set}
\label{sec:uncertainty_set}

As illustrated in the flowchart in Fig.~\ref{fig:schematic}, the first step in our denoising pipeline is the construction of uncertainty sets for the observed phase-space coordinates and their corresponding orbital frequencies.
We constructed the uncertainty set of the observables and orbital quantities in the same manner as in \cite{Hattori2023ApJ...946...48H}. Below we briefly summarize the procedure and highlight the differences specific to the present study.

\subsection{Uncertainty Set of the Observed Quantities}

For each of the $N=518$ stars in our Helmi stream sample, we generate $M=101$ synthetic realizations of the six-dimensional observables,
\eq{
D_i^\mathrm{obs} = \{ (\alpha, \delta, \varpi, \mualpha, \mudelta, \vlos)_{i,j} \mid -50 \le j \le 50 \},
}
representing the observational uncertainties.
The $j=0$ instance corresponds to the nominal point estimate from Gaia DR3. 
To account for the fact that distance uncertainty dominates the errors in phase-space coordinates and orbital frequencies, we specifically define the $j$th realization of the parallax as
\eq{ \label{eq:parallax_sampling}
\varpi_{i,j} = \varpi_i + \left( \frac{j}{25} \right) \sigma_{\varpi,i},
}
where $\varpi_i$ and $\sigma_{\varpi,i}$ are the (zero-point corrected) observed parallax and its uncertainty, respectively. 
This sampling covers a range of $\pm 2\sigma_{\varpi,i}$ in steps of $0.04\sigma_{\varpi,i}$. 
For each $j$, the remaining velocity components $(\mualpha, \mudelta, \vlos)$ are drawn from their respective Gaussian error distributions using the covariance information provided in Gaia DR3. Following \cite{Hattori2023ApJ...946...48H}, we neglect the tiny positional uncertainties in sky coordinates $(\mathrm{RA}, \mathrm{Dec})$ and ignore the covariance between parallax and proper motion, as both have a negligible impact on the recovered frequency distributions.

\subsection{Uncertainty Set of the Orbital Frequency}
\label{sec:Frequency_Uncertainty_Set}

Each element of $D_i^\mathrm{obs}$ is converted into the corresponding Galactocentric position and velocity, $(\vector{x},\vector{v})_{i,j}$, adopting the solar position and motion described in Appendix~\ref{appendix:coordinate}. 
Using these phase-space coordinates, we compute the 
angular momentum $(L_\perp, L_z)$ as well as the orbital energy $E$ 
under the Milky Way potential model of \cite{McMillan2017}, with the \texttt{AGAMA} package \citep{Vasiliev2019_AGAMA}.
We also evaluate the orbital frequencies in the radial, azimuthal, and vertical directions,
\eq{
\vector{\Omega}_{i,j} = (\Omega_R, \Omega_\phi, \Omega_z)_{i,j},
}
for each realization by employing the publicly available code \texttt{naif} \citep{Beraldo2023ascl.soft03004B, Beraldo2023ApJ...955...38B}, which is a Python implementation of the well-established frequency analysis codes \citep{Laskar1990Icar...88..266L, Laskar1993PhyD...67..257L, Valluri1998ApJ...506..686V, Valluri1999ASPC..182..178V, Valluri2010MNRAS.403..525V, Valluri2016ApJ...818..141V}. 
The resulting uncertainty sets of orbital frequencies,
\eq{
D_i^\mathrm{freq} = \{ \vector{\Omega}_{i,j} \mid -50 \leq j \leq 50 \},
}
serve as the inputs for the Greedy Optimistic Clustering analysis described in Section \ref{sec:GOEM_freq}.
By design, the index $j=0$ corresponds to the point estimate of the orbital frequency derived from the observed data, while the remaining indices represent realizations sampled from the observational error distribution. 

This procedure allows us to propagate the measurement uncertainties of Gaia DR3 into the orbital parameter space in a consistent manner, enabling a clustering analysis of the 518 candidate Helmi-stream stars.

\begin{figure*}
\centering 
\includegraphics[width=0.98\textwidth ]{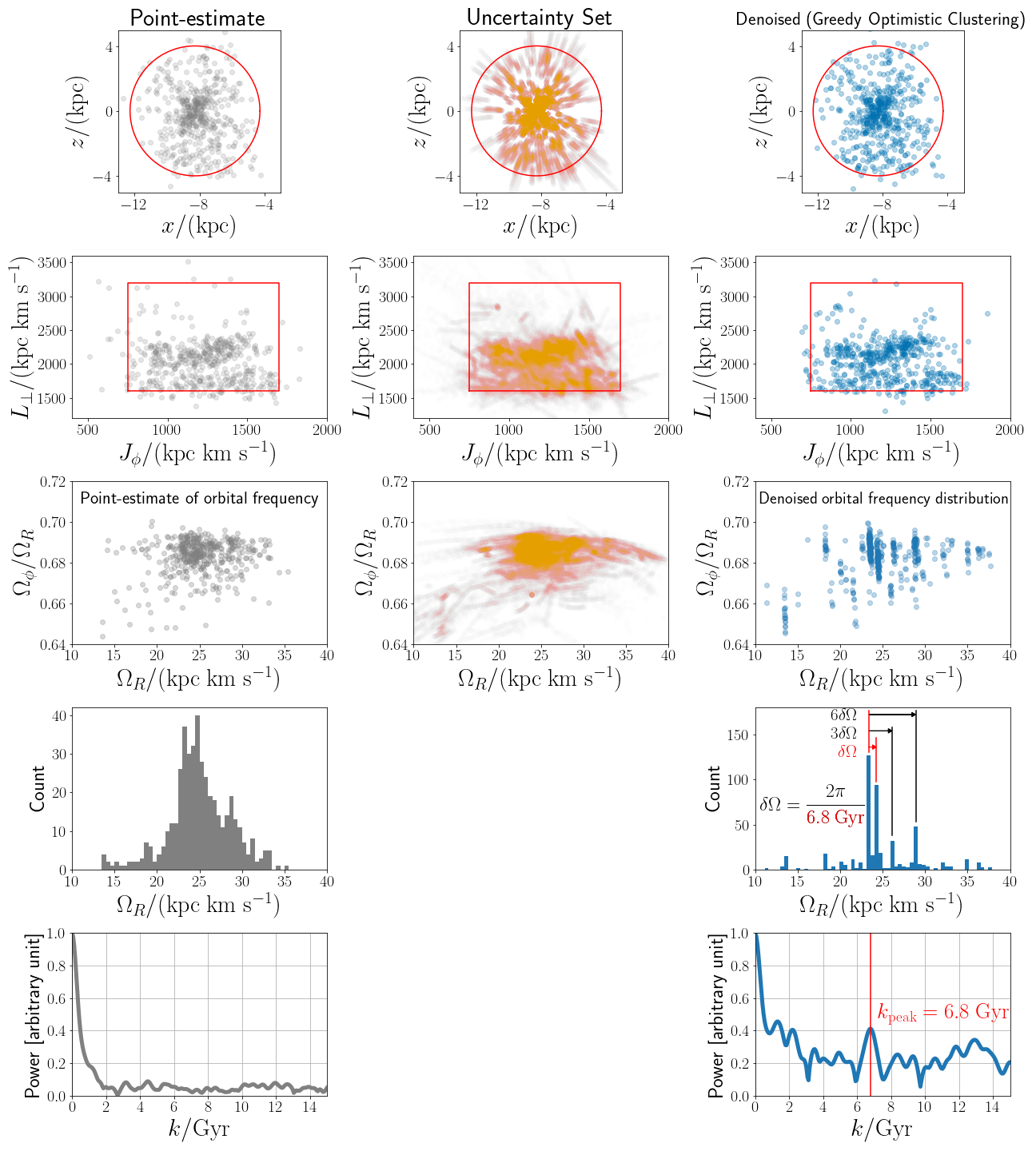}
\caption{
Comprehensive overview of the data selection and the analysis pipeline. The columns illustrate the transition from point-estimates (left) through the construction of uncertainty sets (middle) to a representative denoised realization (right). Top row: Spatial distribution in the $(x, z)$ plane. The red circle indicates $d = 4\,\mathrm{kpc}$. The middle panel shows the uncertainty sets, which are elongated due to heliocentric distance errors. Second row: Angular momentum distribution. The rectangular box indicates `Box B' in \citet{Koppelman2019A&A...625A...5K}, where most Helmi stream stars reside. Third row: Frequency-space distribution. The initially blurred frequency distribution (left) is reorganized into a clumpy structure (right) by the Greedy Optimistic Clustering algorithm (Section~\ref{sec:GOEM_freq}). Fourth row: Histograms of $\Omega_R$. The denoised realization (right) exhibits discrete peaks, whose relative spacings are integer multiples of $\delta \Omega = 2\pi/(6.8\,\mathrm{Gyr})$. Fifth row: The 1-dimensional power spectra derived from the $(\Omega_R, \Omega_\phi)$ distributions. The horizontal axis $k$ represents the trial accretion time in units of Gyr. The emergence of a clear peak at $k_\mathrm{peak} \simeq 6.8\,\mathrm{Gyr}$ in the denoised spectrum (blue) is consistent with the periodic spacing observed in the denoised histogram, indicating that our pipeline provides an objective way of estimating the dynamical age of the stream (Section~\ref{sec:FT_analysis}). 
}
\label{fig:summary_denoise_all}
\end{figure*}

\section{Greedy Optimistic Clustering in Orbital-Frequency Space} \label{sec:GOEM_freq}

The core objective of this study is to resolve the discrete clumps of the Helmi stream stars in the frequency space that are otherwise hidden by observational noise (see the third block in Fig.~\ref{fig:schematic}). This transformation is visually summarized in Fig.~\ref{fig:summary_denoise_all}, where the raw, blurred Gaia DR3 data (left column) is denoised into high-density clumps (right column). In this section, we describe the Greedy Optimistic Clustering algorithm used to achieve this reconstruction.

We model the intrinsic distribution of orbital frequencies
$\vector{\Omega}=(\Omega_R,\Omega_\phi,\Omega_z)$
with a $C$-component isotropic Gaussian mixture of common dispersion $\sigma_\Omega^2$:
\eq{\label{eq:GMM_freq_brief}
P(\vector{\Omega}\mid\theta)=
\sum_{c=1}^{C}\pi_c 
\mathcal{N} \left(\vector{\Omega}\mid
\langle\vector{\Omega}\rangle_c, \; \sigma_\Omega^2\vector{I}\right),
}
where $\theta=\{(\pi_c,\langle\vector{\Omega}\rangle_c) \mid c=1, \cdots, C \}$,
$\pi_c\ge0$, and $\sum_c\pi_c=1$. 
Here, $\mathcal{N}(\cdot \mid \vector{m}, \vector{s})$ 
denotes a multivariate Gaussian distribution with mean vector $\vector{m}$ and covariance matrix $\vector{s}$, while $\vector{I}$ denotes the $3 \times 3$ identity matrix.
To maintain sensitivity to small-scale structures while ensuring algorithmic convergence, we adopt a fixed dispersion of $\sigma_\Omega = 0.1 \kms\kpc^{-1}$ and $C=200$ throughout this study. 
If one ignores measurement errors and uses point estimates
$\widehat{\vector{\Omega}}_i$ for each star, the conventional Expectation-Maximization algorithm maximizes
\eq{
\ln L=\sum_{i=1}^{N}\ln
\left[\sum_{c=1}^{C}\pi_c \; 
\mathcal{N} \left(\widehat{\vector{\Omega}}_i \mid
\langle\vector{\Omega}\rangle_c, \; \sigma_\Omega^2\vector{I}\right)\right],
}
but this approach fails when frequency uncertainties are large.
This failure is visually evident in the left column of Fig.~\ref{fig:summary_denoise_all}, where the raw point-estimates of the Helmi stream stars appear as a continuous, blurred distribution, making any potential substructure impossible to resolve.

To address this issue, we apply the Greedy Optimistic Clustering algorithm \citep{OkunoHattori2025, Hattori2023ApJ...946...48H} to the uncertainty sets $\{ D_i^{\mathrm{freq}} \mid i=1, \dots, N \}$ constructed in Section~\ref{sec:uncertainty_set}. This algorithm is specifically designed to identify clumps in a data set blurred by observational errors. The method assumes that the true orbital frequencies are tightly clustered and allows each star $i$ to select a single realization $\vector{\Omega}_{i,\beta_i} \in D_i^{\mathrm{freq}}$ that best enhances the global clustering. Following the terminology in \citet{OkunoHattori2025}, we refer to this selected realization as the representative point (or denoised value) for star $i$. 
As demonstrated in the right column of Fig.~\ref{fig:summary_denoise_all}, these representative points collectively transform the noisy $\Omega$-distribution into a denoised $\Omega$-distribution with a clumpy structure.

In this algorithm, we jointly estimate the mixture parameters and these discrete per-star indices $\beta_i \in \{-50, \dots, 50\}$ by maximizing the penalized log-likelihood function:
\eq{\label{eq:GO_objective_brief}
f&=\sum_{i=1}^{N}\ln
\left[\sum_{c=1}^{C}\pi_c \; 
\mathcal{N} \left(\vector{\Omega}_{i,\beta_i}\mid
\langle\vector{\Omega}\rangle_c, \; \sigma_\Omega^2\vector{I}\right)\right] \nonumber \\
&-\sum_{i=1}^{N}\pen(i,\beta_i,\lambda),
}
where the penalty term $\pen$ is defined as 
\eq{
\pen(i, \beta_i,\lambda) = 
\frac{1}{2} \lambda \left( \frac{\varpi_{i, \beta_i} - \varpi_i}{\sigma_{\varpi,i}} \right)^2.
}
Here, $\varpi_{i, \beta_i}$ is the parallax value associated with the chosen representative point, and $\lambda$ controls how far each realization may deviate from its observed value $\varpi_i$. Using the definition of our sampling in Equation~\eqref{eq:parallax_sampling}, the penalty simplifies to $\frac{1}{2} \lambda (\beta_i / 25)^2$. The limit $\lambda \to \infty$ reproduces the standard GMM based on point estimates, whereas $\lambda = 0$ treats all realizations equally. Following experiments similar to those in \cite{OkunoHattori2025}, we adopt $\lambda=10^{-4}$ in our fiducial analysis (see also Section~\ref{sec:validation} for the justification of our choice of $\lambda$).

Optimization proceeds in an iterative manner, as in \cite{Hattori2023ApJ...946...48H}, by alternately updating (1) the indices $\{\beta_i\}$ of the representative points and (2) the cluster centroids and weights until the solution converges. 
During optimization, we also employ the split-and-merge procedure of \cite{Ueda1998_NIPS1998_253f7b5d} to improve the solution. 
We adopt $N_{\rm real}=400$ initial conditions for this optimization process, varying the initial choice of the representative indices $\{\beta_i\}$ and the initial locations of the centroids. 
These runs yield similarly good solutions, and their spread represents the uncertainty in our final result.
As we describe in Section~\ref{sec:FT_analysis}, we use these 400 solutions to derive the dynamical age of the Helmi stream.

\section{Fourier Analysis of the Frequency--Space Distribution}
\label{sec:FT_analysis}

Following the successful recovery of a semi-regular lattice of clumps (as shown in the right column of Fig.~\ref{fig:summary_denoise_all}), we now quantify the periodicity of this distribution to estimate the dynamical age of the Helmi stream.
Here we employ a Fourier-based method, broadly following the formalism originally proposed by \citet{Gomez2010MNRAS.401.2285G}. In our framework, this analysis is not performed on a single point-estimate distribution, but on the ensemble of denoised realizations obtained from the Greedy Optimistic Clustering analysis.

\subsection{Ensemble of denoised frequency--space realizations}

Because of observational uncertainties and the highly non--convex structure of the likelihood surface associated with the Greedy Optimistic Clustering analysis, multiple local optima with comparable likelihood values exist. To explore this degeneracy, we perform the clustering analysis using $N_{\rm real}=400$ initial conditions as described in Section~\ref{sec:GOEM_freq}, varying the starting configurations of the cluster centroids and representative points. Each run $s$ converges to a specific set of optimal indices $\{\beta_i\}^{(s)}$, yielding a denoised set of orbital frequencies
\eq{
\bigl\{
\vector{\Omega}_i^{(s)} 
\mid i=1,\dots,N 
\bigr\},
}
where $i$ indexes stars and $s=1,\ldots,N_{\rm real}$ labels the realization. This notation $\vector{\Omega}_i^{(s)}$ corresponds to the representative point $\vector{\Omega}_{i, \beta_i}$ optimized during the $s$th run (see Section~\ref{sec:GOEM_freq}). Note that for a given star $i$, the chosen index $\beta_i$ (and thus the resulting frequency) may vary across realizations $s$ depending on the initial conditions of the clustering optimization. 
While this implies that the denoised frequency of any individual star is not uniquely determined, we treat the ensemble of realizations as a statistical representation of the underlying density structure. Our mock data analysis suggests that while individual realizations may occasionally exhibit spurious features, a good fraction of the ensemble consistently recovers the common clumpy structure required for a reliable dynamical age estimation.

All $N_{\rm real}$ realizations are treated symmetrically as plausible
representations of the underlying frequency--space structure of the Helmi
stream.  The Fourier analysis described below is applied independently to each
realization.

\subsection{Binning of the frequency--space distribution}

For a given realization $s$, we consider the two--dimensional distribution of
stars in $(\Omega_R,\Omega_\phi)$ space.  Following
\citet{Gomez2010MNRAS.401.2285G}, we discretize this distribution by binning the
data on a uniform $N_{\mathrm{grid}}\times N_{\mathrm{grid}}$ grid with bin size
$\Delta$ in each direction. 
Specifically, we define the grid over the range $0 \kms\kpc^{-1} \leq \Omega \leq 100 \kms\kpc^{-1}$ in each coordinate, with $N_\mathrm{grid}=1000$ bins yielding a resolution of $\Delta=0.1 \kms\kpc^{-1}$. 
Let
\begin{equation}
h^{(s)}(m_R,m_\phi)
\end{equation}
denote the number of stars in the cell indexed by $(m_R,m_\phi)$, where
\begin{equation}
m_R,m_\phi = 0,1,\ldots,N_{\mathrm{grid}}-1.
\end{equation}
The array $h^{(s)}(m_R,m_\phi)$ thus represents a discretized image of the
frequency--space distribution for the $s$th realization.

\subsection{Two--dimensional discrete Fourier transform}

For each realization $s$, we compute the two--dimensional discrete Fourier transform of $h^{(s)}(m_R,m_\phi)$ as
\eq{
& H^{(s)}(k_R,k_\phi) = \nonumber \\
& \sum_{m_R=0}^{N_{\mathrm{grid}}-1}
\sum_{m_\phi=0}^{N_{\mathrm{grid}}-1}
h^{(s)}(m_R,m_\phi) 
\exp \left[
-2\pi i \frac{k_R m_R + k_\phi m_\phi}{N_{\mathrm{grid}}}
\right],
}
where $N_{\mathrm{grid}}=1000$ and the integer wavenumbers
\begin{equation}
k_R, k_\phi = -\frac{N_{\mathrm{grid}}}{2}, \ldots, \frac{N_{\mathrm{grid}}}{2}-1
\end{equation}
label the discrete Fourier modes.

The squared modulus $|H^{(s)}(k_R,k_\phi)|^2$ represents the Fourier power at wavenumber $(k_R,k_\phi)$. In practice, the transform is evaluated using a Fast Fourier Transform. For numerical convenience, the binned image and the resulting Fourier array are reordered such that the zero--frequency component $(k_R,k_\phi)=(0,0)$ is located at the center of the array; this choice affects only the indexing convention and does not alter the Fourier amplitudes or the resulting power spectrum.

\subsection{One--dimensional power spectra}

To identify the dominant periodicity, we extract the power along the principal axes. Following our numerical implementation, we define a one-dimensional summary statistic $P^{(s)}(k)$ as the geometric mean of the magnitudes along the $k_R$ and $k_\phi$ axes:
\begin{equation}
P^{(s)}(k) = \sqrt{ |H^{(s)}(k, 0)| \cdot |H^{(s)}(0, k)| },
\end{equation}
where $k \ge 0$. This highlights features that are simultaneously periodic in both frequency coordinates.
While $P^{(s)}(k)$ is technically the square root of the power, we refer to it as `power' hereafter for brevity.

\subsection{Ensemble statistics and accretion time}

At each wavenumber $k$, the ensemble of realizations yields a set of values $\{P^{(s)}(k) \mid s=1,\dots, N_{\rm real} \}$. We summarize the distribution of $P^{(s)}(k)$ by computing the 2.5, 16, 50, 84, and 97.5 percentile values across the ensemble at fixed $k$. Connecting these percentile points as a function of $k$ defines the median power spectrum and the associated uncertainty bands.

While the ensemble spread reflects the numerical variance of the Greedy Optimistic Clustering optimization, our fiducial result is derived from the median power spectrum of the ensemble generated with $\lambda=10^{-4}$. We identify the dominant peak of the median spectrum as a function of $k$ and interpret its location as the characteristic Fourier scale of the frequency--space structure. For tidally disrupted debris accreted at time $T_\mathrm{accretion}$ in the past, the characteristic separation of adjacent clumps in frequency space scales approximately as
\begin{equation}
\delta\Omega \simeq \frac{2\pi}{T_\mathrm{accretion}},
\end{equation}
which implies that a periodic pattern produces a peak in the Fourier spectrum at
\begin{equation}
k_\mathrm{peak} \simeq T_\mathrm{accretion}.
\end{equation}

\begin{figure}
\centering 
\includegraphics[width=0.45\textwidth ]{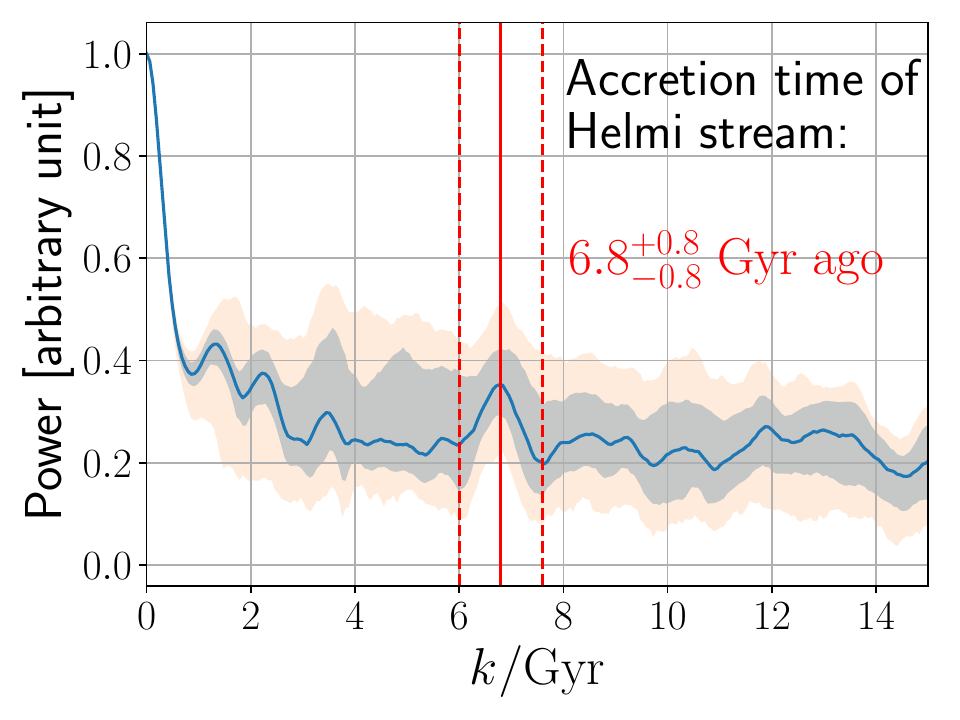}
\caption{
Fourier power spectra of the frequency-space distribution of Helmi-stream stars. 
The blue solid line indicates the median power spectrum derived from an ensemble of 400 denoised $\Omega$-distribution solutions. 
The pale blue and pale orange shaded regions represent the $1\sigma$ (16--84th percentiles) and $2\sigma$ (2.5--97.5th percentiles) uncertainty bands, respectively, reflecting the variance across the denoising realizations. 
The primary peak, representing the characteristic scale used to estimate the dynamical age $T_{\mathrm{accretion}}$, is identified at $k_{\mathrm{peak}} = 6.8$ Gyr (red vertical solid line). 
The uncertainty in this location is defined by the full spectral support of the peak, bounded by the adjacent local minima (dips) at $k_{\mathrm{low}} = 6.0$ Gyr and $k_{\mathrm{high}} = 7.6$ Gyr (red vertical dashed lines).
}
\label{fig:power_spectrum_Gaia}
\end{figure}

\begin{figure}
\centering 
\includegraphics[width=0.35\textwidth ]{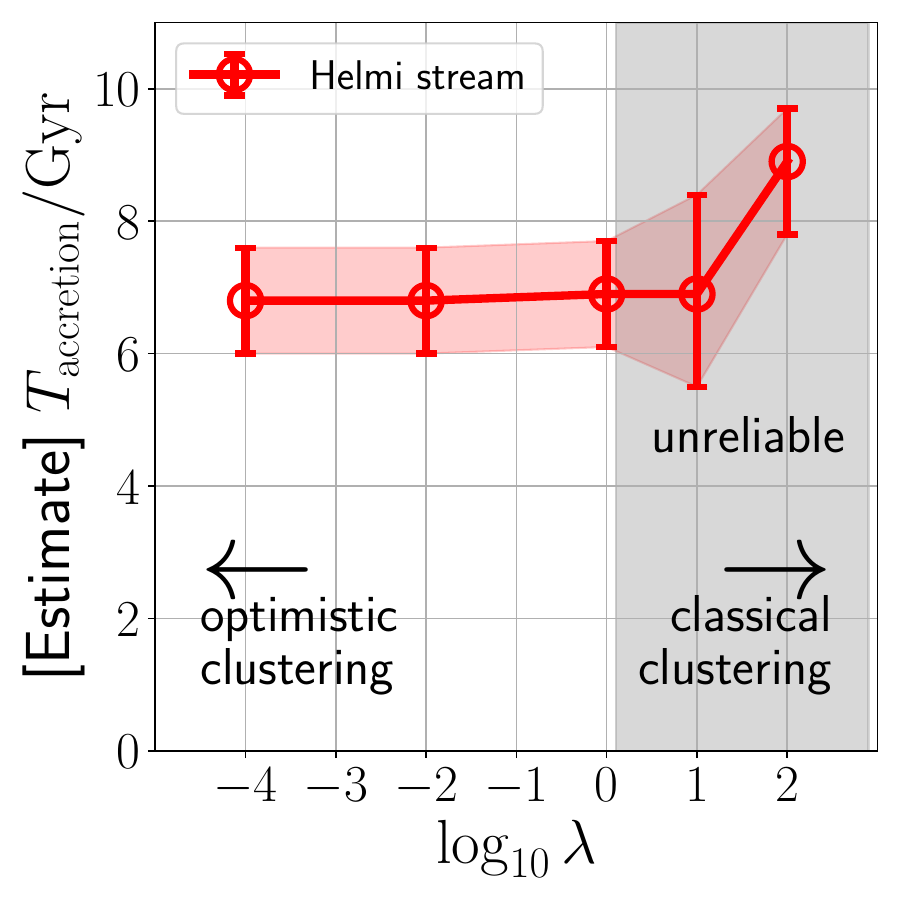}
\caption{
Estimated accretion time $T_{\mathrm{accretion}}$ as a function of the penalty hyperparameter $\lambda$. The points and error bars represent the central estimate and the dip-to-dip uncertainty interval, respectively. Our results are stable across the range $10^{-4} \leq \lambda \leq 1$, yielding an age of $6.8 \pm 0.8$ Gyr. The shaded gray region ($\lambda > 1$) indicates the regime where the Greedy Optimistic Clustering results become less reliable; in this high-penalty limit, the stellar distribution is forced toward the point-estimate values, where measurement uncertainties blur the density contrast and suppress the periodic signal required for the Fourier analysis. 
}
\label{fig:Gaia_lambda}
\end{figure}

As shown in Fig.~\ref{fig:power_spectrum_Gaia}, we identify the primary peak for the Helmi stream at $k_\mathrm{peak} = 6.8\;\mathrm{Gyr}$ (solid vertical line). To quantify the uncertainty in this estimate, we adopt a conservative approach based on the spectral support of the signal. We identify the local minima (or ``dips'') immediately adjacent to the primary peak, denoted as $k_\mathrm{low}$ and $k_\mathrm{high}$ (dashed vertical lines). For the Helmi stream, these occur at $k_\mathrm{low} = 6.0\;\mathrm{Gyr}$ and $k_\mathrm{high} = 7.6\;\mathrm{Gyr}$, respectively. We therefore report the final estimate as $k_\mathrm{peak} = 6.8^{+0.8}_{-0.8}\;\mathrm{Gyr}$, where the uncertainty range corresponds to the full dip-to-dip interval $[k_\mathrm{low}, k_\mathrm{high}]$. This choice ensures that our uncertainty bounds encompass the likely range of the physical signal. This implies a dynamical age (accretion time) of the Helmi stream
\eq{
T_\mathrm{accretion} = k_\mathrm{peak} = 6.8^{+0.8}_{-0.8}\;\mathrm{Gyr}.
}
To assess the reliability of our age estimate, we investigate the dependence of $T_{\mathrm{accretion}}$ on the penalty hyperparameter $\lambda$. We find that our estimate of $T_{\mathrm{accretion}} = 6.8 \pm 0.8$ Gyr remains stable for $10^{-4} \leq \lambda \leq 1$. While adopting $\lambda = 10$ maintains nearly the same central value with a slightly larger uncertainty ($6.9 \pm 1.5$ Gyr), results with $\lambda > 1$ are generally treated with caution. 
As discussed in Section~\ref{sec:validation}, our mock data tests indicate that the algorithm becomes less reliable in this high-penalty regime. This is likely because as $\lambda \to \infty$, the algorithm is forced to fit the distribution of the point estimates, which are significantly blurred by measurement uncertainties. This blurring washes out the density contrast between adjacent clumps, effectively masking the underlying periodic structure.
We therefore adopt $\lambda = 10^{-4}$ for our fiducial analysis (see Fig.~\ref{fig:Gaia_lambda}).

We further validate our methodology using mock data in Section~\ref{sec:validation}. 

\section{Validation with test-particle simulations}
\label{sec:validation}

Having derived the dynamical age for the observed Helmi stream, we now validate our methodology using controlled test-particle simulations. These simulations allow us to verify that the combination of Greedy Optimistic Clustering \citep{OkunoHattori2025} and the Fourier-space analysis \citep{Gomez2010MNRAS.401.2285G} can successfully recover the dynamical age of a disrupted system when the ground truth is known. By applying the same pipeline to error-added mock data, we demonstrate that our framework remains useful in the presence of the Gaia-like observational uncertainties.

We perform four simulations that mimic the tidal disruption of the Helmi stream progenitor, with true accretion ages of $T_{\mathrm{accreted}}^\mathrm{true} = 4, 6, 8,$ and $10$ Gyr ago. Each simulation generates a mock stellar stream whose dynamical evolution is governed by the specified $T_{\mathrm{accreted}}^\mathrm{true}$, enabling a direct comparison between the input and recovered values.

\subsection{Initial conditions and progenitor model}

As the present-day ($t=0$) phase-space coordinates of the progenitor, we adopt those of a core Helmi stream member identified by \cite{Koppelman2019A&A...625A...5K}. This star is assumed to represent the current location of the center of the disrupted system. For each simulation, we integrate the progenitor orbit backward in time for a duration corresponding to $T_{\mathrm{accreted}}^\mathrm{true}$ within a static Milky Way potential.

At the onset of the simulation ($t = -T_{\mathrm{accreted}}^\mathrm{true}$), we generate $N_{\mathrm{particle}} = 10^6$ test particles around the progenitor by adding Gaussian position and velocity offsets with dispersions of $\sigma_x = 1$\,kpc and $\sigma_v = 20 \kms$, respectively. These values are chosen to approximate the spatial extent and internal velocity dispersion of a typical dwarf galaxy progenitor; however, the precise values are not critical for our validation, as they primarily govern the initial spread of stars in frequency space rather than the evolution of the characteristic frequency spacing.

\subsection{Mock data construction}
The orbits of all particles are integrated forward to the present day ($t = 0$), forming extended debris streams that exhibit multiple wraps around the Galaxy. To transform these simulated particles into a realistic Gaia-like dataset, we first perturb the true phase-space coordinates of all mock stars with Gaussian noise. The noise amplitudes for astrometry and radial velocities are assigned based on the Gaia DR3 performance model. From these perturbed data, we derive point estimates for parallaxes ($\varpi$), proper motions, and line-of-sight velocities. Next, we select a solar-neighborhood sample based on these observed quantities. To mimic our main analysis, we select stars that satisfy two criteria: (1) the $2\sigma$ parallax interval must be within 4 kpc of the Sun (i.e., $\varpi + 2\sigma_\varpi > 0.25$ mas), and (2) the parallax must be determined with a signal-to-noise ratio greater than two ($\varpi/\sigma_\varpi > 2$). For each mock realization, both the total number of selected particles and the distribution of point-estimated parallaxes are kept approximately the same as those of the real Helmi-stream sample, ensuring a consistent comparison.

\subsection{Recovery of Accretion Time}

Each simulated dataset is analyzed using the same pipeline applied to the real Gaia data. We adopt $\lambda = 10^{-4}$ as our fiducial penalty hyperparameter for the Greedy Optimistic Clustering, following the methodology established in our previous work \citep{Hattori2023ApJ...946...48H, OkunoHattori2025}. To verify the stability of this choice, we perform a sensitivity analysis by varying $\lambda$ over six orders of magnitude ($10^{-4} \leq \lambda \leq 10^2$).

\begin{figure*}
\centering 
\includegraphics[width=0.45\textwidth ]{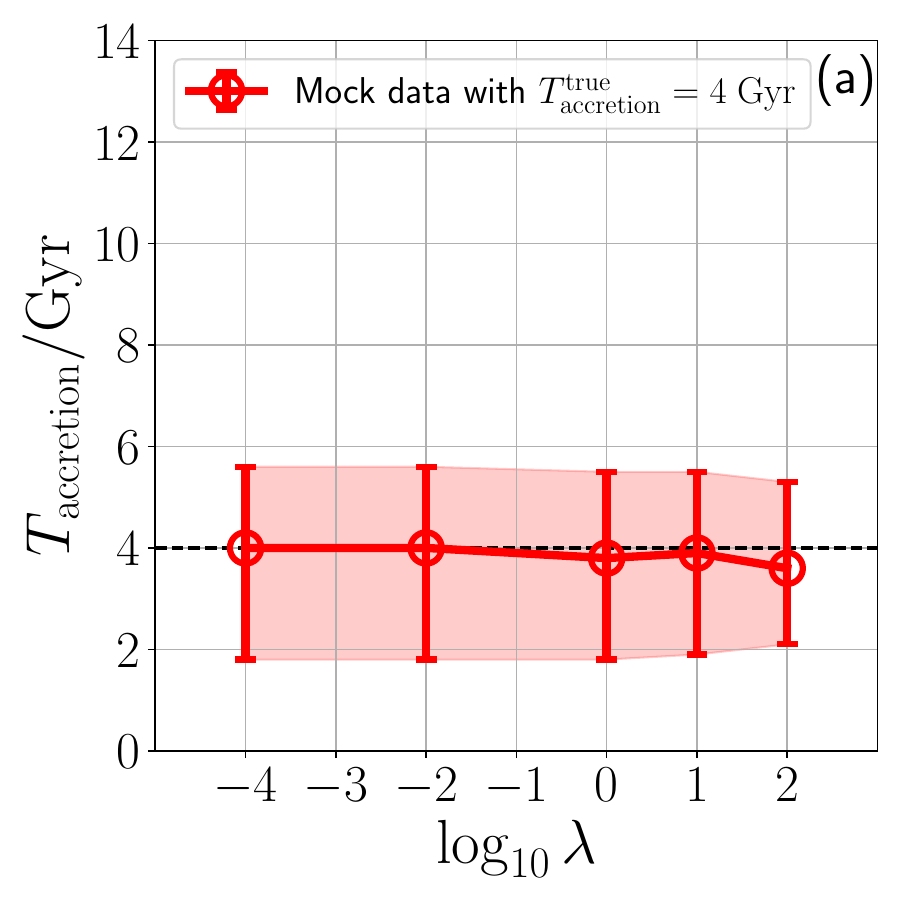}
\includegraphics[width=0.45\textwidth ]{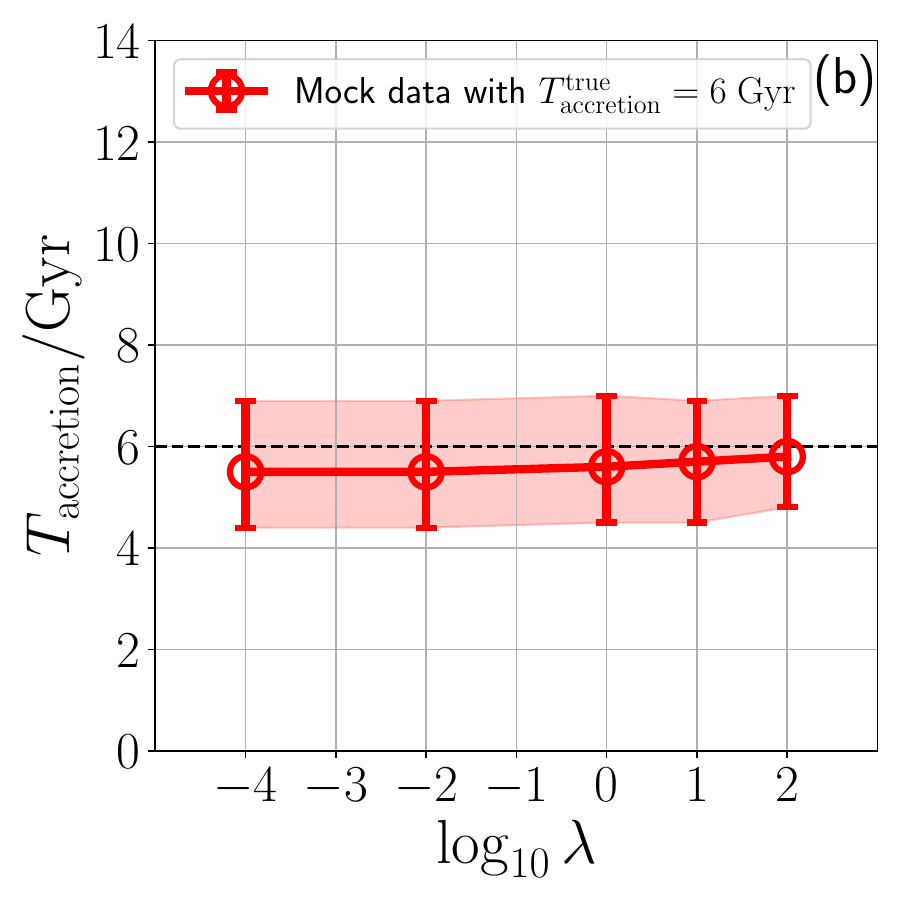}
\includegraphics[width=0.45\textwidth ]{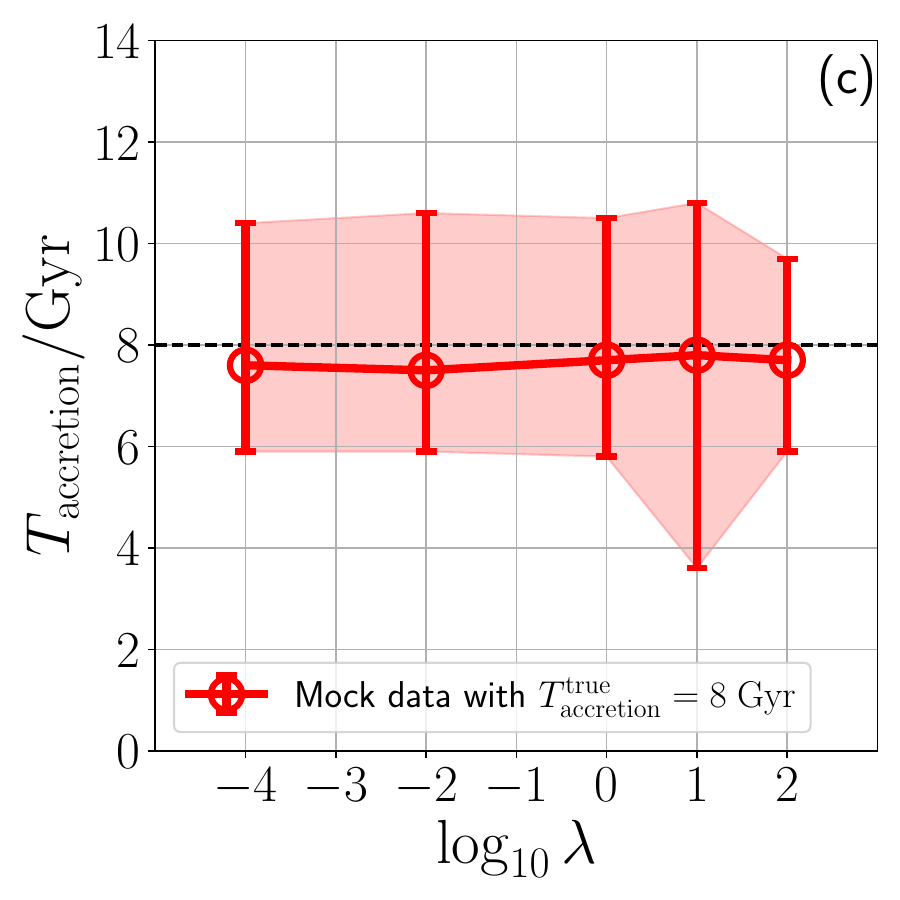}
\includegraphics[width=0.45\textwidth ]{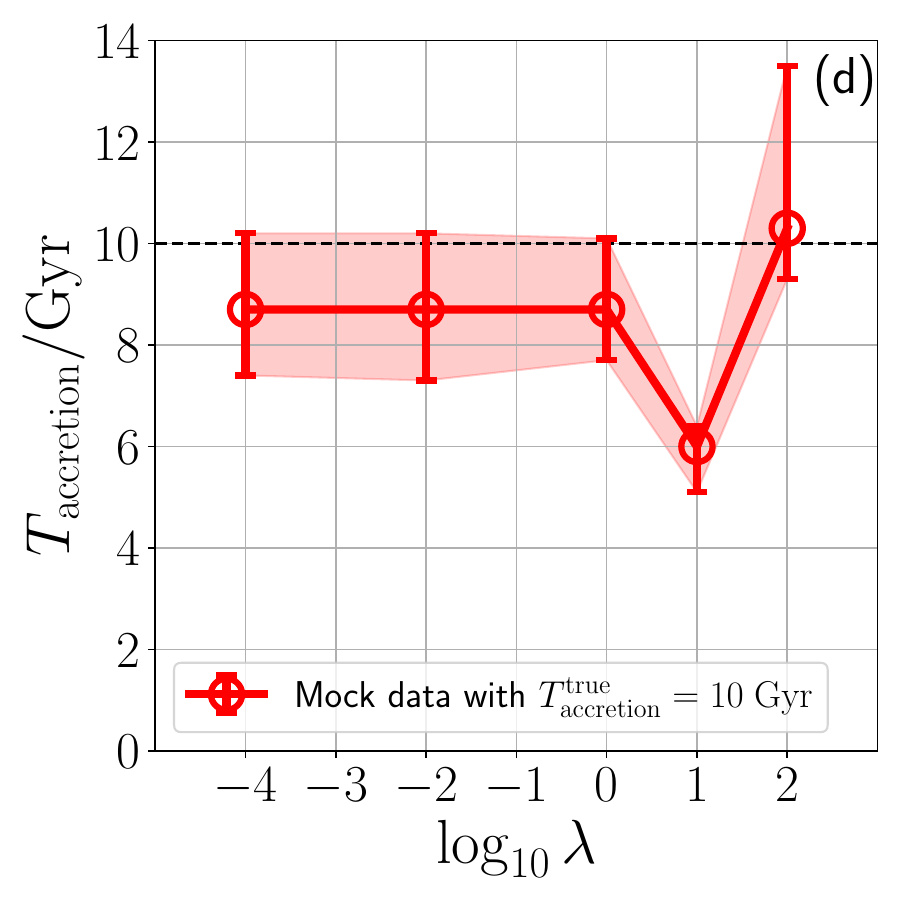}
\caption{
Recovered accretion time as a function of the hyperparameter $\lambda$ across four mock simulations. 
Panels (a), (b), (c), and (d) correspond to true accretion times of $T_\mathrm{accretion}^\mathrm{true} = 4, 6, 8,$ and $10$ Gyr, respectively. 
In each panel, the vertical axis indicates the estimated $T_\mathrm{accretion}$ inferred from the primary peak of the median power spectrum. 
The error bars represent the ``dip-to-dip'' uncertainty range, and the horizontal dashed line marks the ground-truth value of $T_\mathrm{accretion}^\mathrm{true}$. 
The horizontal axis shows the penalty hyperparameter $\lambda$ on a logarithmic scale. 
In the limit of $\lambda \to \infty$, the frequency-space distribution strictly corresponds to the $(\Omega_R, \Omega_\phi)$ values derived from the as-observed astrometric data. 
In contrast, smaller values of $\lambda$ allow the algorithm to explore the observational uncertainty sets to maximize the density contrast of the frequency-space distribution. 
In other words, setting $\lambda \to 0$ effectively sharpens the individual clumps by reducing their internal scatter, thereby revealing the periodic structures that are otherwise smeared by observational noise. 
We observe a clear stability plateau for $\lambda \leq 1$, where the recovered accretion times and their associated uncertainty intervals remain nearly invariant across four orders of magnitude, justifying our fiducial choice of $\lambda = 10^{-4}$.
}
\label{fig:mock_lambda_vs_T}
\end{figure*}

As illustrated in Fig.~\ref{fig:mock_lambda_vs_T}, the estimated $T_{\mathrm{accretion}}$ derived from the primary peak of the median power spectrum agrees closely with the true input values ($T_{\mathrm{accretion}}^\mathrm{true}$) across the range from 4 to 10 Gyr. Our results demonstrate a wide stability plateau for $\lambda \leq 1$, where the recovered accretion ages remain nearly identical despite the four-order-of-magnitude change in the penalty weight. This invariance indicates that in the low-to-moderate penalty regime, the recovered signal is driven primarily by the intrinsic frequency-space structure of the Helmi stream rather than the specific weight of the penalty term.

Conversely, we find that the recovery becomes less stable when adopting higher penalty weights ($\lambda > 1$). To understand this behavior, consider a star in our sample with an observed parallax $\varpi_i \pm \sigma_{\varpi}$. When $\lambda$ is large, the cost of adopting a parallax value significantly different from the observed value (point-estimate) $\varpi_i$ becomes prohibitive due to the quadratic nature of the penalty function. In this regime, the Greedy Optimistic Clustering algorithm is effectively discouraged from exploring the uncertainty set, essentially ``locking'' stars near their as-observed frequencies $(\Omega_R, \Omega_\phi)$. This suppresses the algorithm's ability to resolve the fine-scale frequency structures associated with older streams, which often require subtle shifts within the observational error bars to reveal the underlying coherence.

\begin{figure*}
\centering 
\includegraphics[width=0.45\textwidth ]{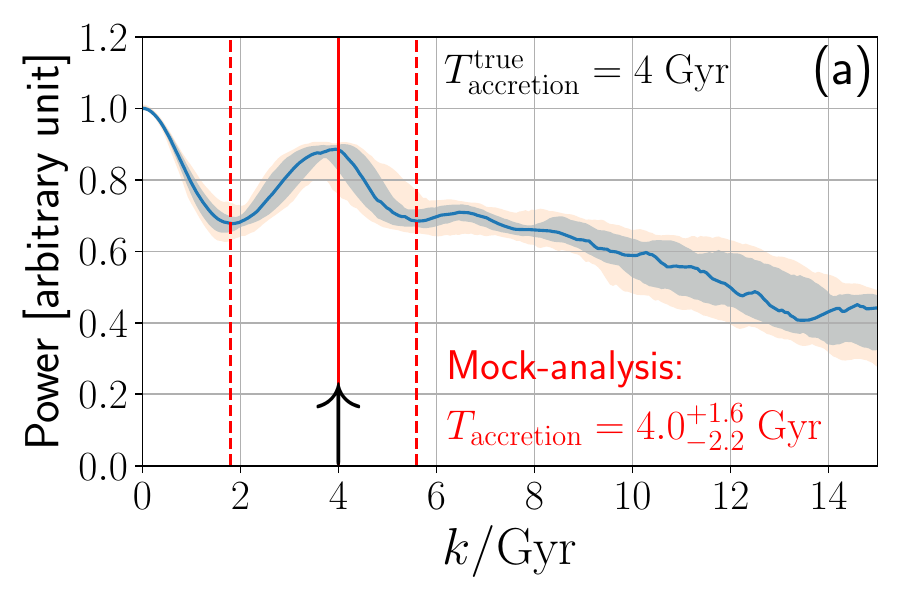}
\includegraphics[width=0.45\textwidth ]{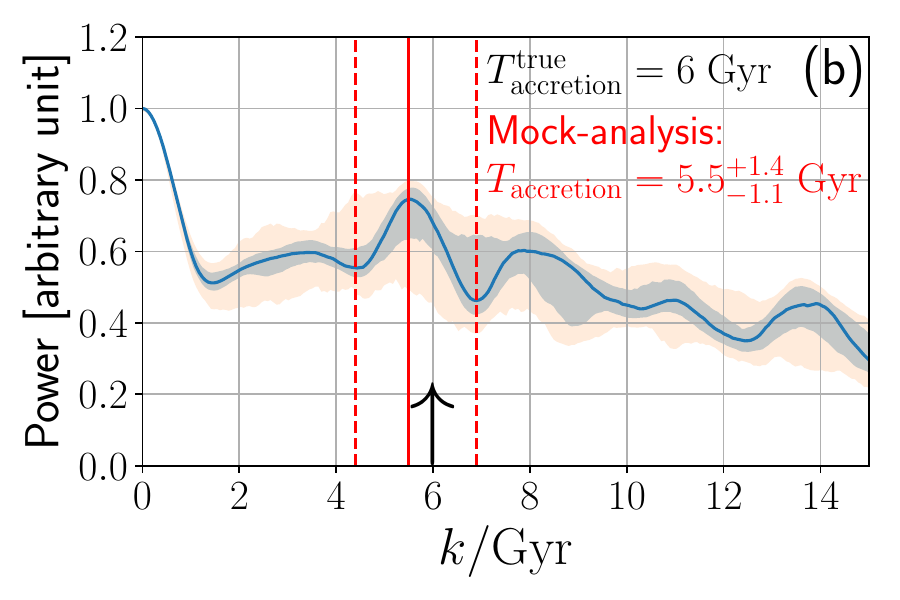}
\includegraphics[width=0.45\textwidth ]{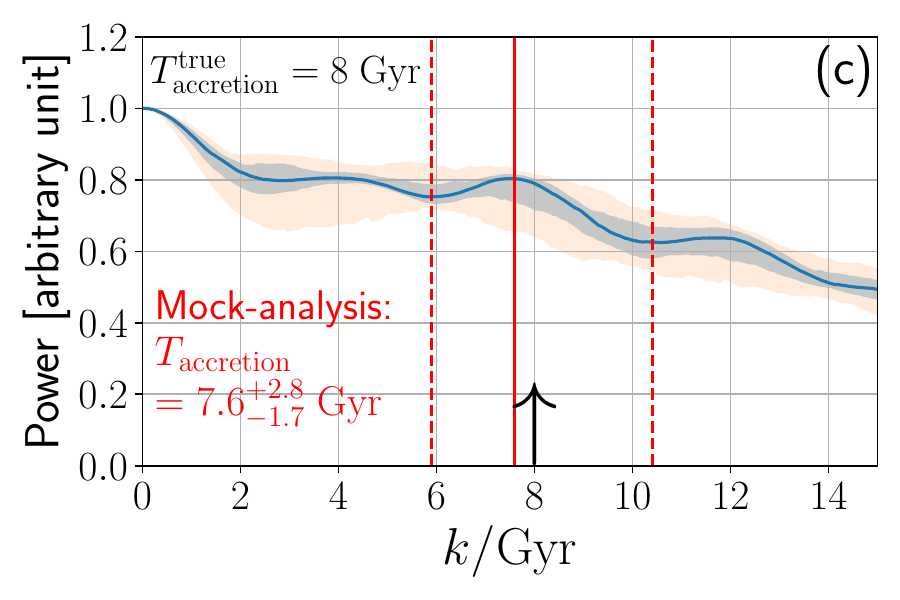}
\includegraphics[width=0.45\textwidth ]{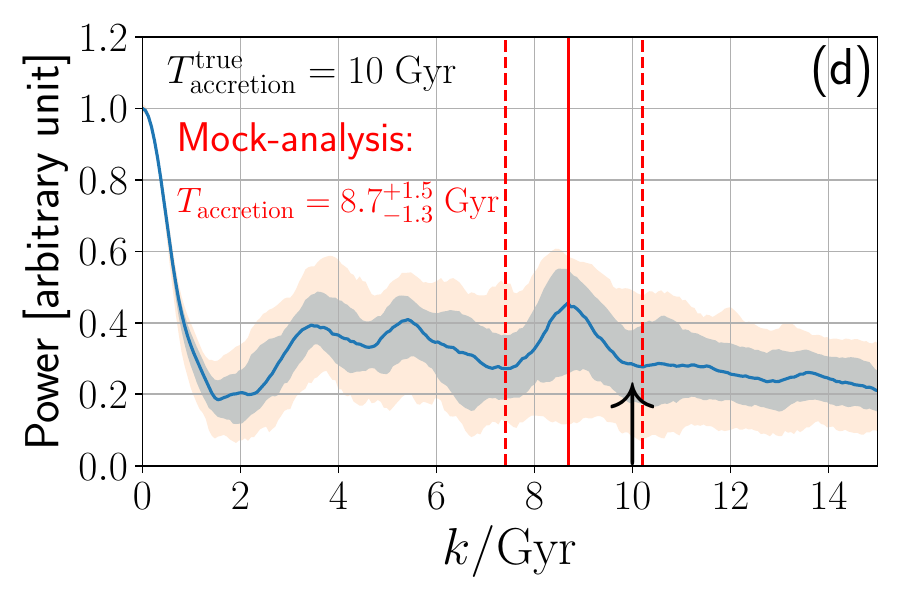}
\caption{
Fourier power spectra of the frequency-space distribution for four mock Helmi-stream simulations with varying dynamical ages. Panels (a), (b), (c), and (d) correspond to true accretion times of $T_\mathrm{accretion}^\mathrm{true} = 4, 6, 8,$ and $10$ Gyr, respectively. In each panel, the power is plotted as a function of wavenumber $k$ (in units of Gyr), such that the primary peak location corresponds to the estimated accretion time. Each mock dataset contains a sample size and observational uncertainty profile equivalent to the Gaia-based Helmi stream sample analyzed in this work. We analyze these data using Greedy Optimistic Clustering with 400 realizations to derive the 2.5, 16, 50, 84, and 97.5 percentile levels of the power distribution at each $k$. The blue solid line indicates the median profile, while the pale blue and pale orange shaded regions represent the $1\sigma$ (16--84th percentiles) and $2\sigma$ (2.5--97.5th percentiles) uncertainty bands, respectively. The peak of the median curve (red vertical solid line) represents our best estimate of the accretion time, with the associated uncertainty defined by the adjacent local minima (or ``dips''; red vertical dashed lines). The ground-truth accretion time $T_\mathrm{accretion}^\mathrm{true}$ is indicated by the black upward arrow. In all four cases, the ``dip-to-dip'' uncertainty range successfully encloses the true accretion time, demonstrating the reliability of our recovery method in the presence of realistic observational noise.
}
\label{fig:mock_power_spectra}
\end{figure*}

To visually inspect the quality of the signal, we show the Fourier power spectra for the four mock simulations in Fig.~\ref{fig:mock_power_spectra}. In all cases, the primary spectral peak approximately corresponds to the true accretion time of the simulation. While these spectral peaks appear marginally more prominent for $\lambda \simeq 1$ in certain cases, the consistent performance across the $\lambda \leq 1$ plateau justifies our fiducial choice of $\lambda = 10^{-4}$.

\begin{figure}
\centering 
\includegraphics[width=0.35\textwidth ]{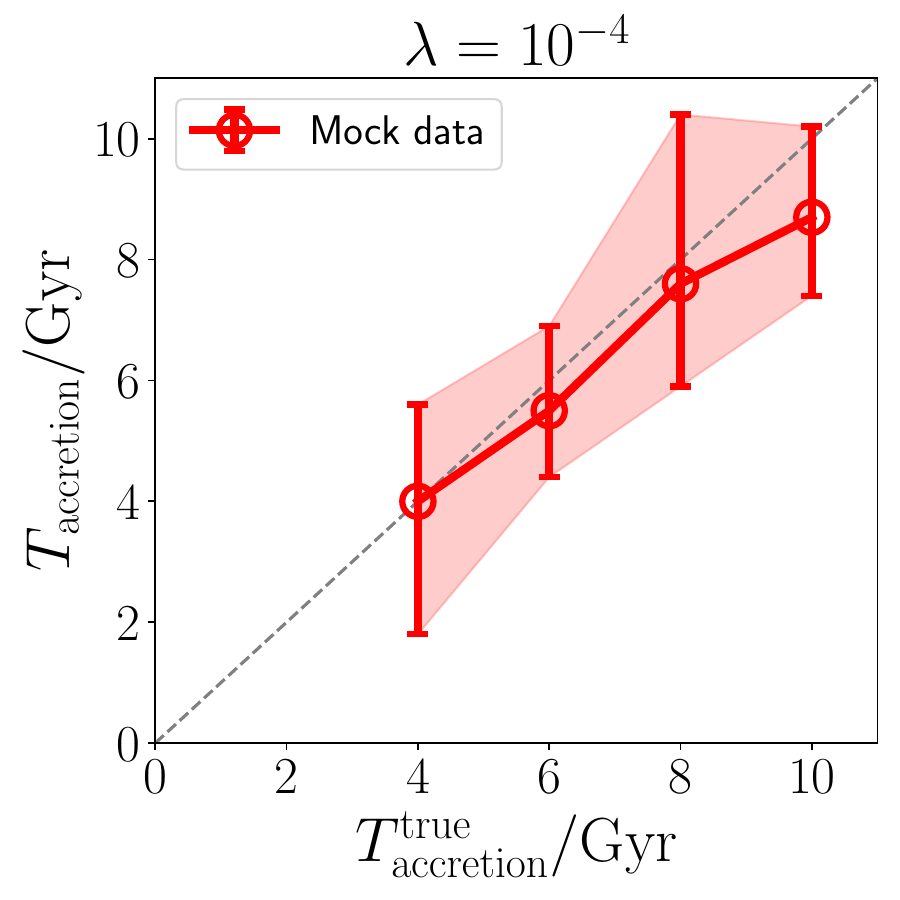}
\caption{
Comparison between the true and recovered accretion times for our four mock Helmi-stream simulations. The horizontal axis represents the ground-truth accretion time ($T_\mathrm{accretion}^\mathrm{true}$) used in the test-particle simulations. The vertical axis shows the recovered estimate ($T_\mathrm{accretion}$) derived by applying our method to the error-added mock data. Data points indicate the location of the primary peak in the median power spectrum ($\lambda = 10^{-4}$), with vertical error bars representing the ``dip-to-dip'' uncertainty interval. The successful recovery of the input ages across the 4--10\,Gyr range demonstrates that Greedy Optimistic Clustering effectively mitigates the smearing effects of observational noise, providing a reliable estimate of the dynamical age.
}
\label{fig:mock_Ttrue_vs_T_lambda_1em4}
\end{figure}

The final validation of our pipeline is summarized in Fig.~\ref{fig:mock_Ttrue_vs_T_lambda_1em4}, which compares the true and recovered ages for our fiducial $\lambda$. In all mock cases, the true value of $T_{\mathrm{accretion}}^\mathrm{true}$ is successfully captured within the ``dip-to-dip'' uncertainty interval. This result validates our use of the full spectral support of the peak as a conservative measure of the dynamical age, even in the presence of realistic Gaia DR3 observational noise.

\subsection{Summary of the mock analysis}

These validation experiments confirm that our framework\hspace{0mm}---which extends the analysis of \cite{Gomez2010MNRAS.401.2285G} by incorporating Greedy Optimistic Clustering---successfully recovers the correct dynamical age of disrupted stellar systems even in the presence of realistic observational noise. By effectively mitigating the smearing of frequency-space structures, this methodology establishes a reliable foundation for determining the accretion history of the Milky Way from observed stellar systems like the Helmi stream.

\section{Discussion and conclusion}
\label{sec:discussion}

\subsection{Summary of Methodology and Results}

In this study, we have determined the dynamical age of the Helmi stream to be $6.8 \pm 0.8$ Gyr. This result was obtained by extending the Fourier-space method of \cite{Gomez2010MNRAS.401.2285G} with a new Greedy Optimistic Clustering framework \citep{OkunoHattori2025}. A limitation of the original frequency-space method is that observational noise smears the substructures needed to identify the characteristic frequency spacing $\delta \Omega$. By allowing for an ``optimistic'' exploration of uncertainty sets, our Greedy Optimistic Clustering algorithm effectively sharpens these frequency-space clumps, revealing a reliable dynamical signal. Extensive validation with error-added mock simulations (Section~\ref{sec:validation}) confirms that this methodology provides a reliable recovery of accretion ages across a wide range of input values ($4$--$10$ Gyr).

\subsection{Implications for the Milky Way's Merger History}

Our derived dynamical age of $\sim 7$ Gyr provides a critical temporal constraint that distinguishes the Helmi stream from other major accretion events. Notably, this accretion epoch is more recent than the massive Gaia-Sausage-Enceladus merger, which is typically dated to $\sim 10$ Gyr ago \citep{Gallart2019NatAs...3..932G, Helmi2020ARAA..58..205H, Montalban2021NatAs...5..640M, Belokurov2023MNRAS.518.6200B} and is responsible for the formation of the bulk of the inner halo (\citealt{Belokurov2018MNRAS.478..611B, Helmi2018Natur.563...85H}; see also \citealt{Chiba2000AJ....119.2843C, Carollo2007Natur.450.1020C, Carollo2010ApJ...712..692C}).

While the stellar populations of the Helmi stream are known to be ancient (more than $\sim 11$ Gyr old; \citealt{Lindsay2025ApJ...989..189L}), our findings indicate that the progenitor system remained a distinct satellite for $\sim 4$ Gyr before being disrupted by the Galactic tidal force. This highlights a fundamental distinction in Galactic archaeology: while stellar ages provide a ``birth certificate'' for the progenitor system, the dynamical age determines its ``arrival date'' into the Milky Way. Our results suggest that the Helmi stream was accreted during a subsequent epoch of Galactic growth, potentially contributing to the vertical heating of a pre-existing proto-disk that was already significant in mass  \citep{Kazantzidis2008ApJ...688..254K, Villalobos2008MNRAS.391.1806V}.

This relatively recent accretion epoch is further supported by the current kinematic state of the Helmi stream. In the solar neighborhood, the Helmi stream stars exhibit a clear asymmetry in their vertical velocities, with a significant majority of stars moving with $v_z < 0$ \citep[e.g.,][]{Helmi1999Natur.402...53H, Kepley2007AJ....134.1579K}. Such a distinct velocity imbalance indicates that the system has not yet reached a symmetric, steady-state distribution in the Galactic potential. Our derived dynamical age of $6.8 \pm 0.8$ Gyr naturally accounts for this observation; a significantly more ancient accretion event would have allowed sufficient time for the orbits to become more evenly distributed, eventually resulting in a more symmetric $v_z$ distribution. Indeed, by comparing this imbalanced $v_z$ distribution with $N$-body simulations, \citet{Koppelman2019A&A...625A...5K} inferred the dynamical age of the Helmi stream to be 5--8 Gyr, a range that is in excellent agreement with our Fourier-based result.

Furthermore, our result is consistent with the star-formation quenching timescale of $\sim 8$ Gyr ago identified by \citet{Ruiz-Lara2022A&A...668L..10R}. Given the $0.8$ Gyr uncertainty in our dynamical age estimate, these two independent findings show a high degree of concordance. This agreement reinforces the conclusion that the progenitor of the Helmi stream survived as a coherent entity for several Gyr before it began being disrupted due to the tidal force from the Milky Way.

\subsection{Limitations and Caveats}

While our methodology provides a reliable estimate of the dynamical age, several physical simplifications should be discussed. First, the recovery of the dynamical age is intrinsically dependent on the assumed Galactic potential. As demonstrated by \cite{Dodd2022A&A...659A..61D}, the frequency-space structure of the Helmi stream is sensitive to the aspherity (e.g., prolate or oblate) of the dark matter halo. While we have adopted the \cite{McMillan2017} Milky Way model, which assumes a spherical dark matter distribution, the precise location of the frequency-space clumps could shift if a different halo geometry or mass profile were assumed.

Secondly, our analysis assumes a static Galactic potential that does not account for the mass growth of the Milky Way \citep{Buist2015A&A...584A.120B, Belokurov2023MNRAS.518.6200B} or the secular evolution of its components over the past $\sim 7$ Gyr \citep{Chiba2001ApJ...549..325C}. While $6.8$ Gyr is sufficiently recent that a steady-state potential serves as a reasonable first-order approximation \citep{Wechsler2002ApJ...568...52W, Hammer2007ApJ...662..322H}, a time-dependent potential could introduce systematic shifts in the recovered accretion time (see also \citealt{Miyoshi2020ApJ...905..109M}). Furthermore, although the LMC has been perturbing the Milky Way \citep{Besla2010, Erkal2019, GaravitoCamargo2019, Koposov2019, Conroy2021Natur.592..534C, Petersen2021NatAs...5..251P, Shipp2021ApJ...923..149S, CorreaMagnus2022MNRAS.511.2610C, Vasiliev2023Galax..11...59V,  Vasiliev2024MNRAS.527..437V}, detailed analyses suggest that the impact of the LMC is notably reduced within $R < 30$ kpc from the Galactic center \citep{Erkal2021MNRAS.506.2677E}. Given that the Helmi stream stars are largely confined to this inner region, the LMC likely has a negligible effect on the relative orbital frequencies used in our age estimation.

Beyond the global evolution of the potential, internal non-axisymmetric features---specifically the Galactic bar---may perturb the orbital frequencies of the Helmi stream stars. Given that these stars occupy a large volume of the stellar halo ($5~\text{kpc} < R < 25~\text{kpc}$) and exhibit large vertical excursions ($|v_z| \sim 250 \kms$ for our solar-neighbor sample), they likely experience the bar's influence as a series of impulsive torques during rapid passages through the Galactic mid-plane \citep[analogous to the Ophiuchus stream;][]{Hattori2016}.

These impulsive kicks in $\vector{\Omega}$ introduce a physical source of dynamical diffusion, potentially blurring the discrete clumps in frequency space over several Gyr (see also \citealt{Woudenberg2025A&A...700A.240W, Dillamore2025MNRAS.542.1331D}). We emphasize that while our Greedy Optimistic Clustering framework effectively mitigates the observational blurring caused by measurement uncertainties, it does not account for this underlying physical scattering. However, as discussed in Appendix~\ref{appendix:bar}, the magnitude of these bar-induced frequency shifts is likely notably smaller than the characteristic frequency gap ($\delta \Omega$) for high-inclination orbits like those of the Helmi stream. The detection of a clear primary peak in our power spectrum (Fig.~\ref{fig:power_spectrum_Gaia}) may hint that the fundamental frequency-space structure has remained coherent enough to allow for a reliable age determination. 

A quantitative assessment of the bar's cumulative effect on the frequency lattice is a significant task reserved for future time-dependent modeling,\footnote{Measuring the degree of this physical scattering within the frequency islands could eventually provide a new method to constrain the strength and evolution of the Galactic bar.} which might give insights into the history of the Galactic bar in the last several Gyr \citep{Hattori2016, PriceWhelan2016ApJ...824..104P, Pearson2017NatAs...1..633P, Dillamore2023MNRAS.524.3596D, Dillamore2024MNRAS.532.4389D, Dillamore2025MNRAS.542.1331D, Woudenberg2025A&A...700A.240W}.

\subsection{Future Prospects and the Utility of Greedy Optimistic Clustering}

The success of the Greedy Optimistic Clustering framework in this study demonstrates that an ``optimistic'' approach to data analysis is a powerful tool for overcoming the current limitations of astrometric uncertainties. By exploring the observational uncertainty sets rather than treating noisy measurements as fixed points, we have shown that it is possible to recover intrinsic dynamical signatures that would otherwise remain hidden.

This approach is not limited to the Helmi stream; in principle, it can be applied to any disrupted stellar system in the Milky Way halo to build a more comprehensive timeline of the Galaxy's assembly history. As the baseline and precision of the Gaia mission continue to grow with future data releases (e.g., DR4 and beyond), the increased accuracy in proper motions and parallaxes will allow the Greedy Optimistic Clustering algorithm to resolve even finer and more ancient structures in frequency space. The synergy between denoising algorithms and high-precision astrometry thus offers a promising path toward a more detailed understanding of the accretion history of the Milky Way.

\acknowledgments

K.H. thanks Masashi Chiba, Leandro Beraldo e Silva, Monica Valluri, Vasily Belokurov, Akifumi Okuno, and Yoshikazu Terada for fruitful discussion. 
K.H. is supported by JSPS KAKENHI Grant Numbers JP24K07101, JP21K13965 and JP21H00053.

This work has made use of data from the European Space Agency (ESA) mission
{\it Gaia} (\url{https://www.cosmos.esa.int/gaia}), processed by the {\it Gaia}
Data Processing and Analysis Consortium (DPAC,
\url{https://www.cosmos.esa.int/web/gaia/dpac/consortium}). Funding for the DPAC
has been provided by national institutions, in particular the institutions
participating in the {\it Gaia} Multilateral Agreement.

\facility{Gaia}

\software{
AGAMA \citep{Vasiliev2019_AGAMA},\;
matplotlib \citep{Hunter2007},
numpy \citep{vanderWalt2011},
scipy \citep{Jones2001}}

\bibliographystyle{aasjournal}
\bibliography{mybibtexfile}

\appendix

\section{Physical Nature of Bar-Induced Perturbations on High-$|v_z|$ Orbits}
\label{appendix:bar}

In this Appendix, we evaluate the potential dynamical influence of the Galactic bar on the orbital frequency distribution of the Helmi stream. A primary concern in any long-term dynamical study is that non-axisymmetric features, such as a rotating bar, may perturb the orbital actions and frequencies, potentially erasing the coherent ``clump'' structures required for our Fourier-based age estimation.

\subsection{The Impulsive Torque Regime}

The impact of the Galactic bar on a stellar stream depends critically on the orbital geometry of its member stars. As demonstrated in numerical simulations of the Ophiuchus stream \citep{Hattori2016}, stars characterized by large vertical excursions experience the bar's gravitational influence as a series of discrete, impulsive torques. Because these stars spend the majority of their orbital period far from the Galactic mid-plane, the torque is only non-negligible during their rapid passages through the disk.

The Helmi stream stars in our sample exhibit a similar dynamical behavior, with characteristic vertical velocities $|v_z| \sim 300 \kms$ near the Galactic mid-plane. At these velocities, the interaction time during a disk crossing is extremely short compared to the radial or azimuthal orbital periods. Consequently, the change in the orbital frequency $\vector{\Omega}$ can be modeled as a kick received at each crossing, rather than a continuous secular drift.

\subsection{Quantitative Assessment of Frequency Scattering}

To evaluate the magnitude of this effect, we perform a back-of-the-envelope calculation using a representative Helmi stream star in our $T_\mathrm{accretion}^\mathrm{true} = 6$~Gyr simulation. As discussed in Section~\ref{sec:showcase} and illustrated in the right-most column of Fig.~\ref{fig:simulation_no_error}, a representative star (e.g., star E) has completed roughly $n_R = 25$ radial cycles since accretion and follows a prograde orbit with $|L_z| \simeq 1000 \kpckms$.

The characteristic frequency gap between adjacent islands in our simulation is approximately
\eq{
\delta \Omega_R = \frac{2\pi}{T_\mathrm{accretion}^\mathrm{true}} = \frac{2\pi}{6\text{ Gyr}} \simeq 1.05 \kmskpc .
}

We estimate the actual physical kick $\Delta \Omega_{\mathrm{kick}}$ as follows. Assuming the star crosses the disk ($|v_z| = 300 \kms$) through an effective thickness of $H \simeq 1 \kpc$ of the Galactic disk, the interaction time is $\Delta t \simeq 3.3$~Myr. Adopting a conservative bar torque of $\tau \simeq 0.5 \kpckms \Myr^{-1}$ at a pericenter of $R_\mathrm{peri} = 5 \kpc$ \citep{Hattori2016}, the change in angular momentum per crossing is $\Delta L_z = \tau \Delta t \simeq 1.65 \kpc\kms$. The resulting frequency kick is:
\eq{
\Delta \Omega_{\mathrm{kick}} \simeq \frac{\Delta L_z}{R_\mathrm{peri}^2} \simeq \frac{1.65 \kpckms}{(5 \kpc)^2} \simeq 0.066 \kmskpc .
}
Assuming that the bar's phase at each disk crossing is uncorrelated, the cumulative physical scattering after $n_R = 25$ radial cycles follows a random walk, 
\eq{
\Delta \Omega_{\mathrm{phys}} \simeq \sqrt{n_R} \Delta \Omega_{\mathrm{kick}} \simeq 0.33 \kmskpc. 
}
This cumulative physical perturbation is roughly three times smaller than the characteristic frequency gap ($\delta \Omega_R \simeq 1.05 \kmskpc$). This scale separation confirms that while the bar could induce a non-negligible broadening of the frequency islands, the discrete structure would remain well-defined against bar-induced dynamical diffusion.

\subsection{Conservation of the Orbital Cycle Count}

The fundamental signal used in this study is the frequency spacing $\delta \Omega$, which arises from the discrete difference in the number of radial cycles ($n_R$ vs. $n_R-1$) completed by stars currently in the solar neighborhood (see Section~\ref{sec:showcase}). For a perturbation to invalidate this ``clock,'' it would need to physically move a star from one frequency island to another---effectively changing its total integer count of oscillations since the merger. 

The calculation above demonstrates that bar-induced kicks are too weak to move a star between frequency islands. Instead, the bar's influence is manifested as a jitter around the island centers, leaving the global orbital history---and thus the timing signal---intact.

\subsection{Summary: Stability Against Galactic Perturbations}

In summary, while the Galactic bar is a major non-axisymmetric feature, its effect on the Helmi stream is minimized by two factors:
\begin{itemize}
\item
Geometric isolation: High-$|v_z|$ orbits minimize the time spent in the region of the bar's strongest non-axisymmetric potential, resulting in an impulsive rather than secular interaction.
\item
Spectral scale separation: The cumulative frequency shifts induced by the bar ($\sim 0.3 \kmskpc$ over $\sim 6$ Gyr) are notably smaller than the frequency spacing characteristic of a 6--7 Gyr accretion event ($\sim 1 \kmskpc$).
\end{itemize}
These results imply that the orbital frequency spacing $\delta \Omega$ is a persistent feature that remains even in a realistic, non-axisymmetric Milky Way. Consequently, our derived dynamical age is reliable, and is largely unaffected by perturbations from the Galactic bar.

\section{Coordinate system} \label{appendix:coordinate}

We adopt a right-handed Galactocentric Cartesian coordinate system $(x,y,z)$, in which the $(x,y)$-plane is the Galactic disk plane. 
The position of the Sun is assumed to be $\vector{x}_\odot = (x_\odot,y_\odot,z_\odot) = (-R_\odot,0,z_\odot)$, with $R_\odot = 8.277 \kpc$ \citep{GRAVITY2022A&A...657L..12G} and $z_\odot = 0.0208 \kpc$ \citep{Bennett2019MNRAS.482.1417B}. 
The velocity of the Sun with respect to the Galactic rest frame is assumed to be $\vector{v}_\odot = (v_{x,\odot},v_{y,\odot},v_{z,\odot}) = (9.3, 251.5, 8.59) \kms$.

\end{document}

%% file: newcommands.tex
\newcommand{\GH}{G\'omez \& Helmi} 

\newcommand{\msun}{\mbox{$M_{\odot}$}}

\def\vector#1{\mbox{\boldmath $#1$}}

\newcommand{\kpc}{\ensuremath{\,\mathrm{kpc}}}
\newcommand{\Myr}{\ensuremath{\,\mathrm{Myr}}}
\newcommand{\Gyr}{\ensuremath{\,\mathrm{Gyr}}}
\newcommand{\kms}{\ensuremath{\,\mathrm{km\ s}^{-1}}}
\newcommand{\kmskpc}{\ensuremath{\,\mathrm{{km\ s}^{-1}\ {kpc}^{-1}}}}
\newcommand{\kpckms}{\ensuremath{\,\mathrm{kpc\;}\mathrm{{km\ s}^{-1}}}}

\newcommand{\vlos}{v_{\ensuremath{\mathrm{los}}}}

\newcommand{\mualpha}{\mu_{\alpha*}}
\newcommand{\mudelta}{\mu_\delta}
\newcommand{\omr}  {\Omega_r}

\newcommand{\eq}[1]{\begin{align}#1\end{align}}

